\documentclass{article}
       \usepackage{cc,times}

\usepackage{url,mathrsfs}
\usepackage{amsmath}
\usepackage{amsfonts}
\usepackage{graphicx}
\usepackage{subfigure}

\begin{document}

\title{\vspace{-0.in}Compressed Counting\vspace{-0.in}}

\author{ Ping Li\\
       Department of Statistical Science\\
       Faculty of Computing and Information Science \\
       Cornell University,        Ithaca, NY 14853\\
       pingli@cornell.edu\vspace{-0.in}}

\maketitle
\vspace{-0.in}
\begin{abstract}
\vspace{-0.09in}
\textbf{Counting} is a fundamental operation.  For example,  counting the $\alpha$th frequency moment,  $F_{(\alpha)} = \sum_{i=1}^D A_t[i]^\alpha$, of a streaming signal $A_t$ (where $t$ denotes time), has been an active area of research, in theoretical computer science, databases, and data mining. When $\alpha =1$, the task (i.e., counting the sum) can be accomplished using a  counter. When $\alpha \neq 1$, however, it becomes non-trivial to design a small space (i.e., low memory) counting system.

{\em Compressed Counting (CC)} is proposed for efficiently  computing the $\alpha$th frequency moment of a data stream  $A_t$, where $0<\alpha\leq2$. CC is applicable if the streaming data follow the {\em Turnstile} model, with the restriction that at the time $t$ for the evaluation, $A_t[i]\geq 0, \forall i\in[1,D]$, which includes the {\em strict Turnstile} model as a special case. For  data streams  in practice, this restriction is  minor.

The underlying technique is  {\em skewed stable random projections}, which  captures the intuition that, when $\alpha =1$ a simple counter suffices, and when $\alpha = 1\pm\Delta$ with small $\Delta$, the sample complexity  should be low (continuously as a function of $\Delta$).  We show the sample complexity (number of projections) {\small$k = G \frac{1}{\epsilon^2}\log\left(\frac{2}{\delta}\right)$}, where {\small$G  =  O\left(\epsilon\right)$} as  {\small$\Delta\rightarrow 0$}. In other words, for small $\Delta$, $k = O\left({1}/{\epsilon}\right)$ instead of $O\left({1}/{\epsilon^2}\right)$.

The case $\Delta\rightarrow 0$ is practically very important. It is now well-understood that one can obtain good approximations to the entropies of data streams using the $\alpha$th moments with $\alpha=1\pm\Delta$ and very small $\Delta$. For statistical inference using the {\em method of moments}, it is sometimes reasonable use the $\alpha$th moments with $\alpha$ very close to 1.  As another example,
$\Delta$ might be the ``decay rate'' or ``interest rate,'' which is usually small. Thus, {\em Compressed Counting} will be an ideal tool, for estimating the total value in the future, taking in account the effect of decaying or interest accruement.

Finally, our another contribution is an algorithm for approximating the logarithmic norm, {\small$\sum_{i=1}^D\log A_t[i]$}, and the logarithmic distance, {\small$\sum_{i=1}^D\log\left|A_t[i] - B_t[i]\right|$}. The logarithmic norm arises in  statistical estimations. The logarithmic distance is  useful in machine learning practice with heavy-tailed data.

\end{abstract}

\vspace{-0.25in}
\section{Introduction}
\vspace{-0.05in}

This paper 
focuses on {\em counting}, which is among the most fundamental operations in almost every field of science and engineering. Computing the sum {\small$\sum_{i=1}^DA_t[i]$} is the simplest counting ($t$ denotes time). Counting the {\small$\alpha$th moment $\sum_{i=1}^DA_t[i]^\alpha$} is more general. When {\small$\alpha\rightarrow 0+$, $\sum_{i=1}^DA_i[i]^\alpha$} counts the total number of non-zeros in $A_t$. When {\small$\alpha = 2$, $\sum_{i=1}^DA_t[i]^\alpha$} counts the ``energy'' or  ``power'' of the signal $A_t$. If  $A_t$ actually outputs the power of an underlying signal $B_t$, counting the sum {\small$\sum_{i=1}^DB_t$} is equivalent to computing {\small$\sum_{i=1}^DA_t[i]^{1/2}$}.

Here, $A_t$ denotes a time-varying signal, for example, {\em  data streams}\cite{Book:Henzinger_99,Proc:Feigenbaum_FOCS99,Article:Indyk_JACM06,Proc:Babcock_PODS02,Article:Indyk_TKDE03,Article:Muthukrishnan_05}. In the literature, the $\alpha$th frequency moment of a data stream $A_t$ is defined as\vspace{-0.05in}
{\small \begin{align}
F_{(\alpha)} = \sum_{i=1}^D|A_t[i]|^\alpha.
 \end{align}}\vspace{-0.15in}

Counting $F_{(\alpha)}$ for massive data streams is practically important, among many  challenging issues in data stream computations. In fact, the general theme of ``scaling up for high dimensional data and high speed data streams'' is among the  ``ten challenging problems in data mining research.''

Because the elements, $A_t[i]$, are time-varying,  a na\'ive counting mechanism requires a system of $D$ counters to compute $F_{(\alpha)}$ exactly. This is not always realistic when $D$ is large and we only need an approximate answer. For example, $D$ may be  $2^{64}$ if $A_t$ records the arrivals of IP addresses. Or, $D$ can be the total number of checking/savings accounts.

\textbf{\em Compressed Counting (CC)} is a new scheme for approximating the $\alpha$th frequency moments of data streams (where $0<\alpha\leq 2$) using low memory. The underlying technique is based on what we call {\em skewed stable random projections}.

\vspace{-0.02in}
\subsection{The Data Models}

We consider the popular {\em Turnstile} data stream model \cite{Article:Muthukrishnan_05}. The input  stream $a_t = (i_t, I_t)$, $i_t\in [1,\ D]$ arriving sequentially describes the underlying signal $A$, meaning $A_t[i_t] = A_{t-1}[i_t] + I_t$. The increment $I_t$ can be either positive (insertion) or negative (deletion). Restricting $I_t\geq 0$ results in the {\em cash register} model. Restricting  $A_t[i]\geq 0$ at all  $t$ (but $I_t$ can still be either positive or negative) results in the {\em strict Turnstile} model, which suffices for describing most (although not all) natural phenomena. For example\cite{Article:Muthukrishnan_05}, in a database, a record can only be deleted if it was previously inserted. Another example is the checking/savings account, which allows deposits/withdrawals  but generally does not allow overdraft.

{\em Compressed Counting (CC)} is applicable when, at the time $t$ for the evaluation, $A_t[i]\geq 0$ for all $i$. This is more flexible than the {\em strict Turnstile} model, which requires  $A_t[i]\geq 0$ at all $t$. In other words, CC is applicable when data streams  are (a) insertion only (i.e., the {\em cash register} model), or (b) always non-negative (i.e., the {\em strict Turnstile} model), or (c) non-negative at check points. We believe our model suffices for describing most natural data streams  in practice.

With the realistic restriction that $A_t[i]\geq 0$ at $t$, the definition of the $\alpha$th frequency moment becomes

{\vspace{-0.2in}\small \begin{align}\label{eqn_def_F2}
F_{(\alpha)} = \sum_{i=1}^DA_t[i]^\alpha;
 \end{align}}\vspace{-0.15in}

\noindent and the case $\alpha = 1$ becomes trivial, because
{\small\begin{align}\vspace{-0.15in}
F_{(1)} = \sum_{i=1}^DA_t[i]= \sum_{s=1}^t I_s
\end{align}}\vspace{-0.1in}

\noindent In other words, for $F_{(1)}$, we need only a simple counter to accumulate all values of increment/decrement $I_t$.

For $\alpha \neq 1$, however, counting (\ref{eqn_def_F2}) is still a non-trivial problem. Intuitively, there should exist an intelligent counting system that performs almost like a simple counter when $\alpha = 1\pm\Delta$ with small $\Delta$. The parameter $\Delta$ may bear a clear physical meaning. For example, $\Delta$ may be the ``decay rate'' or ``interest rate,'' which is usually small.

The proposed {\em Compressed Counting (CC)} provides such an intelligent counting systems. Because its underlying technique is based on {\em skewed stable random projections}, we provide a brief introduction to {\em skewed stable distributions}.

\vspace{-0.02in}
\subsection{Skewed Stable Distributions}

A random variable $Z$ follows a $\beta$-skewed $\alpha$-stable distribution if the Fourier transform of its density  is\cite{Book:Zolotarev_86}
{\small\begin{align}\notag\vspace{-0.1in}
{\mathscr{F}}_Z(t) &= \text{E}\exp\left(\sqrt{-1}Zt\right)  \hspace{0.3in} \alpha \neq 1,\\\notag
&= \exp\left(-F|t|^\alpha\left(1-\sqrt{-1}\beta\text{sign}(t)\tan\left(\frac{\pi\alpha}{2}\right)\right)\right),
\end{align}}\vspace{-0.1in}

\noindent where $-1\leq \beta\leq 1$ and $F>0$ is the scale parameter. We denote $Z \sim S(\alpha,\beta,F)$. Here $0<\alpha \leq 2$. When $\alpha<0$, the inverse Fourier transform is unbounded; and when $\alpha>2$, the inverse Fourier transform is not a probability density. This is why {\em Compressed Counting} is limited to $0<\alpha\leq2$.

Consider two independent variables, $Z_1, Z_2 \sim S(\alpha, \beta,1)$. For any non-negative constants $C_1$ and $C_2$, the ``$\alpha$-stability'' follows from  properties of Fourier transforms:
{\small\begin{align}\notag\vspace{-0.1in}
Z = C_1Z_1 + C_2Z_2 \sim S\left(\alpha,\beta, C_1^\alpha + C_2^\alpha\right).
\end{align}}
\noindent However, if $C_1$ and $C_2$ do not have the same signs,  the above ``stability'' does not hold (unless $\beta = 0$ or $\alpha = 2$, $0+$). To see this, we consider $Z = C_1 Z_1 - C_2 Z_2$, with $C_1\geq 0$ and $C_2\geq 0$. Then, because $\mathscr{F}_{-Z_2}(t) = \mathscr{F}_{Z_2}(-t) $,
{\small\begin{align}\notag\vspace{-0.1in}
\mathscr{F}_Z =& \exp\left(-|C_1t|^\alpha\left(1-\sqrt{-1}\beta\text{sign}(t)\tan\left(\frac{\pi\alpha}{2}\right)\right)\right) \\\notag \times&\exp\left(-|C_2t|^\alpha\left(1+\sqrt{-1}\beta\text{sign}(t)\tan\left(\frac{\pi\alpha}{2}\right)\right)\right),
\end{align}}\vspace{-0.1in}

\noindent which does not represent a stable law, unless $\beta = 0$ or $\alpha =2$, $0+$. This is the fundamental reason why {\em Compressed Counting} needs the restriction that at the time of evaluation, elements in the data streams should have the same signs.

\vspace{-0.05in}
\subsection{Skewed Stable Random Projections}

Given  $R\in\mathbb{R}^{D}$ with each element $r_i\sim S(\alpha, \beta, 1)$ i.i.d., then\vspace{-0.06in}
{\small\begin{align}\notag
R^\text{T} A_t = \sum_{i=1}^D r_{i} A_t[i] \sim S\left(\alpha, \beta, F_{(\alpha)} = \sum_{i=1}^D A_t[i]^\alpha\right),
\end{align}}\vspace{-0.09in}

\noindent meaning  $R^\text{T}A_t$ represents one sample of the stable distribution whose scale parameter $F_{(\alpha)}$ is  what we are after.

Of course, we need more than one sample to  estimate $F_{(\alpha)}$. We can generate a matrix $\mathbf{R}\in\mathbb{R}^{D\times k}$ with each entry $r_{ij} \sim S(\alpha, \beta, 1)$. The resultant vector $X = \mathbf{R}^\text{T}A_t\in\mathbb{R}^k$ contains $k$ i.i.d. samples: $x_j \sim S\left(\alpha,\beta,F_{(\alpha)}\right)$, $j = 1$ to $k$.

Note that this is a linear projection; and recall that the {\em Turnstile} model is also linear. Thus, {\em skewed stable random projections} can be applicable to dynamic data streams. For every incoming $a_t = (i_t, I_t)$, we update $x_j \leftarrow x_j + r_{i_tj} I_t$ for $j = 1$ to $k$. This way, at any time $t$, we maintain $k$ i.i.d. stable samples. The remaining task is to recover $F_{(\alpha)}$, which is a statistical estimation problem.

\vspace{-0.05in}
\subsection{Counting in Statistical/Learning Applications}

The {\em method of moments} is often convenient and popular in statistical parameter estimation. Consider, for example, the three-parameter generalized gamma distribution $GG(\theta,\gamma,\eta)$, which is highly flexible for modeling positive data, e.g., \cite{Article:Li_SINR06}. If $X\sim GG(\theta,\gamma,\eta)$, then the first three moments are $\text{E}(X) = \theta\gamma$, $\text{Var}(X^2) = \theta\gamma^2$, $\text{E}\left(X - \text{E}(X)\right)^3 = (\eta+1)\theta\gamma^3$. Thus, one can estimate $\theta$, $\gamma$ and $\eta$ from $D$ i.i.d. samples $x_i\sim GG(\theta,\gamma,\eta)$ by counting the first three empirical moments from the data. However, some moments may be (much) easier to compute than others if the data $x_i$'s are collected from data streams. Instead of using integer moments, the parameters can also be estimated from any three {\em fractional} moments, i.e., $\sum_{i=1}^D x_i^\alpha$, for three different values of $\alpha$. Because  $D$ is very large, any consistent estimator is likely to provide a good estimate. Thus, it might be reasonable to choose $\alpha$ mainly based on the computational cost. See Appendix \ref{app_moments} for comments on the situation in which one may also care about the relative accuracy caused by different choices of $\alpha$.

The logarithmic norm $\sum_{i=1}^D\log x_i$ arises in statistical estimation, for example, the maximum likelihood estimators for the Pareto and gamma distributions. Since it is closely connected to the moment problem, Section \ref{sec_log} provides an algorithm for approximating the logarithmic norm, as well as for the logarithmic distance; the latter can be quite useful in machine learning practice with massive heavy-tailed data (either dynamic or static) in lieu of the usual $l_2$ distance.

Entropy is also an important summary statistic. Recently \cite{Proc:Zhao_IMC07} proposed to approximate the entropy moment $\sum_{i=1}^D x_i\log x_i$ using the $\alpha$th moments with $\alpha = 1\pm\Delta$ and very small $\Delta$.

\vspace{-0.05in}
\subsection{Comparisons with Previous Studies}

Pioneered by\cite{Proc:Alon_STOC96}, there have been many studies on approximating the $\alpha$th frequency moment $F_{(\alpha)}$.  \cite{Proc:Alon_STOC96} considered integer moments, $\alpha = 0$, 1, 2, as well as $\alpha>2$. Soon after, \cite{Proc:Feigenbaum_FOCS99,Proc:Indyk_FOCS00} provided improved algorithms for $0<\alpha\leq 2$.
 \cite{Proc:Saks_STOC02,Proc:Kumar_FOCS02} proved the sample complexity lower bounds for $\alpha >2$. \cite{Proc:Woodruff_SODA04} proved the optimal lower bounds for all frequency moments, except for $\alpha = 1$, because  for non-negative data, $F_{(1)}$ can be computed essentially error-free with a counter\cite{Article:Morris_CACM78,Article:Flajolet_BIT85,Proc:Alon_STOC96}.
 \cite{Proc:Indyk_STOC05} provided algorithms for $\alpha >2$ to (essentially) achieve the lower bounds proved in \cite{Proc:Saks_STOC02,Proc:Kumar_FOCS02}.

Note that an algorithm, which ``achieves the optimal bound,'' is not necessarily practical because the constant may be very large. In a sense, the method based on {\em symmetric stable random projections}\cite{Article:Indyk_JACM06} is one of the few successful algorithms that are simple and free of large constants.  \cite{Article:Indyk_JACM06} described the procedure for approximating $F_{(1)}$ in data streams and proved the bound for $\alpha =1$ (although not explicitly). For $\alpha \neq 1$, \cite{Article:Indyk_JACM06} provided a conceptual algorithm.  \cite{Proc:Li_SODA08} proposed various estimators for {\em symmetric stable random projections} and provided the constants explicitly for all $0<\alpha\leq 2$.

None of the previous studies, however, captures of the intuition that, when $\alpha = 1$, a simple counter suffices for computing $F_{(1)}$ (essentially) error-free, and when $\alpha = 1 \pm \Delta$ with small $\Delta$, the sample complexity (number of projections, $k$) should be low and vary continuously as a function of $\Delta$.

\textbf{\em Compressed Counting (CC)} is proposed for $0<\alpha \leq 2$ and  it works particularly well when $\alpha = 1\pm\Delta$  with small $\Delta$.  This can be practically very useful. For example, $\Delta$ may be  the ``decay rate'' or the ``interest rate,'' which is usually small; thus CC  can count the total value in the future taking into account the effect of decaying or interest accruement. In parameter estimations using the {\em method of moments}, one may choose the $\alpha$th moments with $\alpha$ close 1. Also,  one can  approximate the entropy moment using the $\alpha$th moments with $\alpha = 1\pm\Delta$ and very small $\Delta$\cite{Proc:Zhao_IMC07}.

Our study has connections to the  Johnson-Lindenstrauss  Lemma\cite{Article:JL84}, which proved $k = O\left(1/\epsilon^2\right)$ at $\alpha = 2$. An analogous bound holds for  $0<\alpha \leq 2$\cite{Article:Indyk_JACM06,Proc:Li_SODA08}. The dependency on $1/\epsilon^2$ may raise  concerns if, say, $\epsilon\leq 0.1$. We will show that CC achieves $k = O(1/\epsilon)$ in the neighborhood of $\alpha = 1$.

\subsection{Two Statistical Estimators}

Recall that {\em Compressed Counting (CC)} boils down to a statistical estimation problem. That is, given
$k$ i.i.d. samples $x_j \sim S\left(\alpha, \beta=1, F_{(\alpha)}\right)$, estimate the scale parameter $F_{(\alpha)}$. Section \ref{sec_gm} will explain why we fix $\beta = 1$.

\begin{figure}[h]\vspace{-0.1in}
\begin{center}
\includegraphics[width = 2. in]{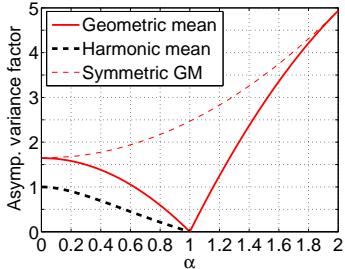}
\end{center}
\vspace{-0.3in}
\caption{Let $\hat{F}$ be an estimator of $F$ with asymptotic variance {\small$\text{Var}\left(\hat{F}\right) = V\frac{F^2}{k} + O\left(\frac{1}{k^2}\right)$}. We plot the $V$ values for the {\em geometric mean} and  the {\em harmonic mean} estimators, along with the $V$ values for the geometric mean estimator in \cite{Proc:Li_SODA08} (symmetric GM). When $\alpha\rightarrow 1$, our method achieves an ``infinite improvement'' in terms of the asymptotic variances.
}\label{fig_comp_var_factor}
\end{figure}

Part of this paper is to provide estimators which are convenient for theoretical analysis, e.g., tail bounds. We provide the {\em geometric mean}  and the {\em harmonic mean} estimators, whose asymptotic variances are illustrated in Figure \ref{fig_comp_var_factor}.

\vspace{-0.1in}

\begin{itemize}
\item {The \textbf{geometric mean} estimator, $\hat{F}_{(\alpha),gm}$}\vspace{-0.05in}
{\small\begin{align}\notag
&\hat{F}_{(\alpha),gm} = \frac{\prod_{j=1}^k |x_j|^{\alpha/k}} { D_{gm}}, \hspace{0.2in} (D_{gm} \ \text{depends on} \  \alpha \ \text{and} \ k).
\\\notag
&\text{Var}\left(\hat{F}_{(\alpha),gm}\right) = \frac{F_{(\alpha)}^2}{k}\frac{\pi^2}{12}\left(\alpha^2+2-3\kappa^2(\alpha)\right)+O\left(\frac{1}{k^2}\right),\\\notag
&\kappa(\alpha) = \alpha, \ \ \ \text{if} \ \  \alpha <1, \hspace{0.2in} \kappa(\alpha) = 2-\alpha, \ \ \ \text{if} \ \  \alpha >1.
\end{align}}
$\hat{F}_{(\alpha),gm}$ is unbiased. We prove the sample complexity explicitly and show  {\small$k = O\left(1/\epsilon\right)$} suffices for $\alpha$ around 1.
\vspace{-0.05in}
\item {The \textbf{harmonic mean} estimator, $\hat{F}_{(\alpha),hm,c}$,  for $\alpha <1$}\vspace{-0.06in}
{\small\begin{align}\notag
&\hat{F}_{(\alpha),hm,c} = \frac{k\frac{\cos\left(\frac{\alpha\pi}{2}\right)}{\Gamma(1+\alpha)}}{\sum_{j=1}^k|x_j|^{-\alpha}}
\left(1- \frac{1}{k}\left(\frac{2\Gamma^2(1+\alpha)}{\Gamma(1+2\alpha)}-1\right) \right), \\\notag
&\text{Var}\left(\hat{F}_{(\alpha),hm,c}\right) = \frac{F^{2}_{(\alpha)}}{k}\left(\frac{2\Gamma^2(1+\alpha)}{\Gamma(1+2\alpha)}-1\right) + O\left(\frac{1}{k^2}\right).
\end{align}}
It is considerably more accurate than $\hat{F}_{(\alpha),gm}$ and its sample complexity bound is also provided in an explicit form.  Here $\Gamma(.)$ is the usual gamma function.
\end{itemize}

\vspace{-0.1in}
\subsection{Paper Organization}

Section \ref{sec_gm} begins with analyzing the moments of skewed stable distributions, from which the {\em geometric mean} and  {\em harmonic mean} estimators are derived. Section \ref{sec_gm} is then devoted to the detailed analysis of the {\em geometric mean} estimator.

Section \ref{sec_hm} analyzes the {\em harmonic mean} estimator. Section \ref{sec_log} addresses the application of CC in statistical parameter estimation and an algorithm for approximating the logarithmic norm and distance. The proofs are presented as appendices.

\vspace{-0.02in}
\section{The Geometric Mean Estimator}\label{sec_gm}
We first prove a fundamental result about the moments of skewed stable distributions.
\begin{lemma}\label{lem_moments}
If $Z \sim S(\alpha,\beta,F_{(\alpha)})$, then for any $-1<\lambda<\alpha$,
{\small\begin{align}\notag
&\textbf{E}\left(|Z|^\lambda\right) = F_{(\alpha)}^{\lambda/\alpha} \cos\left(\frac{\lambda}{\alpha}\tan^{-1}\left(\beta\tan\left(\frac{\alpha\pi}{2}\right)\right)\right) \\\notag
&\times\left(1+\beta^2\tan^2\left(\frac{\alpha\pi}{2}\right)\right)^{\frac{\lambda}{2\alpha}}
\left(\frac{2}{\pi}\sin\left(\frac{\pi}{2}\lambda\right)\Gamma\left(1-\frac{\lambda}{\alpha}\right)\Gamma\left(\lambda\right)\right),
\end{align}}
which can be simplified when $\beta = 1$, to be
{\small\begin{align}\notag
&\textbf{E}\left(|Z|^\lambda\right) = \\\notag
&{F}_{(\alpha)}^{\lambda/\alpha} \frac{\cos\left(\frac{\kappa(\alpha)}{\alpha}\frac{\lambda\pi}{2}\right)}
{\cos^{\lambda/\alpha}\left(\frac{\kappa(\alpha)\pi}{2}\right)}\left(\frac{2}{\pi}\sin\left(\frac{\pi}{2}\lambda\right)\Gamma\left(1-\frac{\lambda}{\alpha}\right)\Gamma\left(\lambda\right)\right),
\\\notag&\kappa(\alpha) = \alpha \ \ \  \text{if}  \ \ \ \alpha<1, \ \  \ \text{and} \  \ \kappa(\alpha)=2-\alpha\ \  \ \text{if} \ \ \alpha>1.
\end{align}}
For $\alpha <1$, and  $-\infty<\lambda <\alpha$,
{\small\begin{align}\notag
\textbf{E}\left(|Z|^\lambda\right) =\textbf{E}\left(Z^\lambda\right) = {F}_{(\alpha)}^{\lambda/\alpha} \frac{ \Gamma\left(1-\frac{\lambda}{\alpha}\right) }
{\cos^{\lambda/\alpha}\left(\frac{\alpha\pi}{2}\right)
\Gamma\left(1-\lambda\right)}.
\end{align}}

\noindent\textbf{Proof:} \hspace{0.2in}
See Appendix \ref{proof_lem_moments}. $\Box$\\\vspace{-0.1in}
\end{lemma}

Recall that {\em Compressed Counting} boils down to estimating $F_{(\alpha)}$ from these $k$ i.i.d. samples $x_j \sim S(\alpha,\beta,F_{(\alpha)})$.  Setting $\lambda = \frac{\alpha}{k}$ in Lemma \ref{lem_moments} yields an unbiased estimator:
{\small\begin{align}\notag
&\hat{F}_{(\alpha),gm,\beta} = \frac{\prod_{j=1}^k |x_j|^{\alpha/k}} { D_{gm,\beta}},\\\notag
&D_{gm,\beta} = \cos^k\left(\frac{1}{k}\tan^{-1}\left(\beta\tan\left(\frac{\alpha\pi}{2}\right)\right)\right) \times\\\notag
& \left(1+\beta^2\tan^2\left(\frac{\alpha\pi}{2}\right)\right)^{\frac{1}{2}}\left[\frac{2}{\pi}\sin\left(\frac{\pi\alpha}{2k}\right)
\Gamma\left(1-\frac{1}{k}\right)\Gamma\left(\frac{\alpha}{k}\right)\right]^k.
\end{align}}

The following Lemma  shows that the variance of $\hat{F}_{(\alpha),gm,\beta}$ decreases with increasing $\beta\in[0,1]$.
\begin{lemma}
The variance of $\hat{F}_{(\alpha),gm,\beta}$
{\small\begin{align}\notag
&\text{Var}\left(\hat{F}_{(\alpha),gm,\beta}\right) = F_{(\alpha)}^2 V_{gm,\beta}\\\notag
&V_{gm,\beta} = \frac{
\cos^k\left(\frac{2}{k}\tan^{-1}\left(\beta\tan\left(\frac{\alpha\pi}{2}\right)\right)\right)}
{\cos^{2k}\left(\frac{1}{k}\tan^{-1}\left(\beta\tan\left(\frac{\alpha\pi}{2}\right)\right)\right)}\times\\\notag
&\hspace{0.5in}\frac{
\left[\frac{2}{\pi}\sin\left(\frac{\pi\alpha}{k}\right)
\Gamma\left(1-\frac{2}{k}\right)\Gamma\left(\frac{2\alpha}{k}\right)\right]^k}
{ \left[\frac{2}{\pi}\sin\left(\frac{\pi\alpha}{2k}\right)
\Gamma\left(1-\frac{1}{k}\right)\Gamma\left(\frac{\alpha}{k}\right)\right]^{2k}}-1,\vspace{-0.1in}
\end{align}}
\noindent is a decreasing function of $\beta \in [0,1]$.

\noindent\textbf{Proof:} \ The result follows from the fact that
{\small\begin{align}\notag
&\frac{\cos\left(\frac{2}{k}\tan^{-1}\left(\beta\tan\left(\frac{\alpha\pi}{2}\right)\right)\right)}
{\cos^2\left(\frac{1}{k}\tan^{-1}\left(\beta\tan\left(\frac{\alpha\pi}{2}\right)\right)\right) } \\\notag
=& 2 - \sec^2\left(\frac{1}{k}\tan^{-1}\left(\beta\tan\left(\frac{\alpha\pi}{2}\right)\right)\right),
\end{align}}\vspace{-0.15in}

\noindent is a deceasing function of $\beta\in[0,1]$.   $\Box$
\end{lemma}

Therefore, for attaining the smallest variance, we take $\beta =1$. For brevity, we simply use $\hat{F}_{(\alpha),gm}$ instead of $\hat{F}_{(\alpha),gm,1}$. In fact, the rest of the paper will always consider $\beta =1$ only.

We rewrite $\hat{F}_{(\alpha),gm}$ (i.e., $\hat{F}_{(\alpha),gm,\beta=1}$) as
{\small\begin{align}\label{eqn_F_gm}
&\hat{F}_{(\alpha),gm} = \frac{\prod_{j=1}^k |x_j|^{\alpha/k}} { D_{gm}}, \hspace{0.5in} (k\geq 2),\\\notag
&D_{gm} = \left(\cos^k\left(\frac{\kappa(\alpha)\pi}{2k}\right)/\cos \left(\frac{\kappa(\alpha)\pi}{2}\right)\right)\\\notag
&\hspace{0.3in}\times  \left[\frac{2}{\pi}\sin\left(\frac{\pi\alpha}{2k}\right)
\Gamma\left(1-\frac{1}{k}\right)\Gamma\left(\frac{\alpha}{k}\right)\right]^k.
\end{align}}
\noindent Here, $\kappa(\alpha) = \alpha$, if $\alpha<1$, and $\kappa(\alpha) = 2-\alpha$ if $\alpha>1$.

Lemma \ref{lem_gm_moments} concerns the asymptotic moments of $\hat{F}_{(\alpha),gm}$.
\begin{lemma}\label{lem_gm_moments}
As $k\rightarrow\infty$
{\small\begin{align}\notag
&\left[\cos\left(\frac{\kappa(\alpha)\pi}{2k}\right)\frac{2}{\pi}\Gamma\left(\frac{\alpha}{k}\right)\Gamma\left(1-\frac{1}{k}\right)\sin\left(\frac{\pi}{2}\frac{\alpha}{k}\right)\right]^k
\\\label{eqn_gm_asymp}
\rightarrow& \exp\left(-\gamma_e\left(\alpha-1\right)\right),
\end{align}}
\noindent \textbf{monotonically} with increasing $k$ ($k\geq 2$), where $\gamma_e  = 0.57724...$ is Euler's constant. \hspace{0.15in}
For any fixed $t$, as $k\rightarrow\infty$,
{\small\begin{align}
&\text{E}\left(\left(
\hat{F}_{(\alpha),gm} \notag
\right)^t\right)\\\notag
 =& F_{(\alpha)}^t\frac{
\cos^{k}\left(\frac{\kappa(\alpha)\pi}{2k}t\right) \left[\frac{2}{\pi}\sin\left(\frac{\pi\alpha}{2k}t\right)
\Gamma\left(1-\frac{t}{k}\right)\Gamma\left(\frac{\alpha}{k}t\right)\right]^{k}}
{
\cos^{kt}\left(\frac{\kappa(\alpha)\pi}{2k}\right) \left[\frac{2}{\pi}\sin\left(\frac{\pi\alpha}{2k}\right)
\Gamma\left(1-\frac{1}{k}\right)\Gamma\left(\frac{\alpha}{k}\right)\right]^{kt}
}\\\notag
=& F_{(\alpha)}^t\exp\left( \frac{1}{k}\frac{\pi^2(t^2-t)}{24}\left(\alpha^2+2-3\kappa^2(\alpha)\right)+O\left(\frac{1}{k^2}\right)\right).
\end{align}}\vspace{-0.1in}
{\small\begin{align}\notag
\text{Var}\left(\hat{F}_{(\alpha),gm}\right) = \frac{F_{(\alpha)}^2}{k}\frac{\pi^2}{12}\left(\alpha^2+2-3\kappa^2(\alpha)\right)+O\left(\frac{1}{k^2}\right).
\end{align}}
\noindent\textbf{Proof:} See Appendix \ref{proof_lem_gm_moments}. $\Box$
\end{lemma}

In (\ref{eqn_F_gm}), the denominator $D_{gm}$ depends on $k$ for small $k$. For convenience in analyzing tail bounds, we consider an asymptotically equivalent {\em geometric mean} estimator:
{\small\begin{align}\notag
\hat{F}_{(\alpha),gm,b} = \exp\left(\gamma_e(\alpha-1)\right)\cos\left(\frac{\kappa(\alpha)\pi}{2}\right)\prod_{j=1}^k |x_j|^{\alpha/k}.
\end{align}}\vspace{-0.1in}

\begin{figure}[h]
\begin{center}
\mbox{
\subfigure[Right  bound, $\alpha<1$]{\includegraphics[width=1.6in]{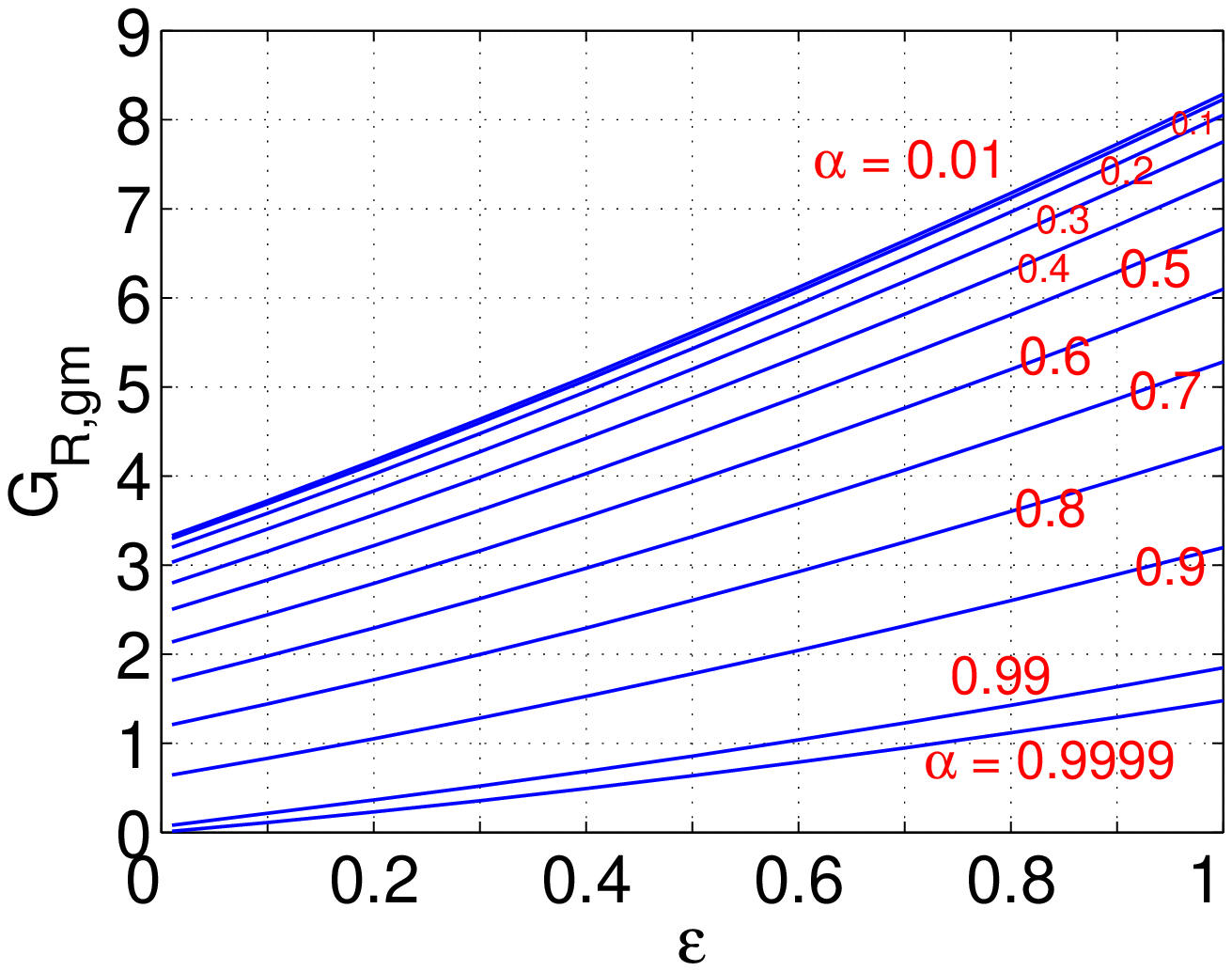}}\hspace{-0.1in}
\subfigure[Right  bound, $\alpha>1$]{\includegraphics[width=1.6in]{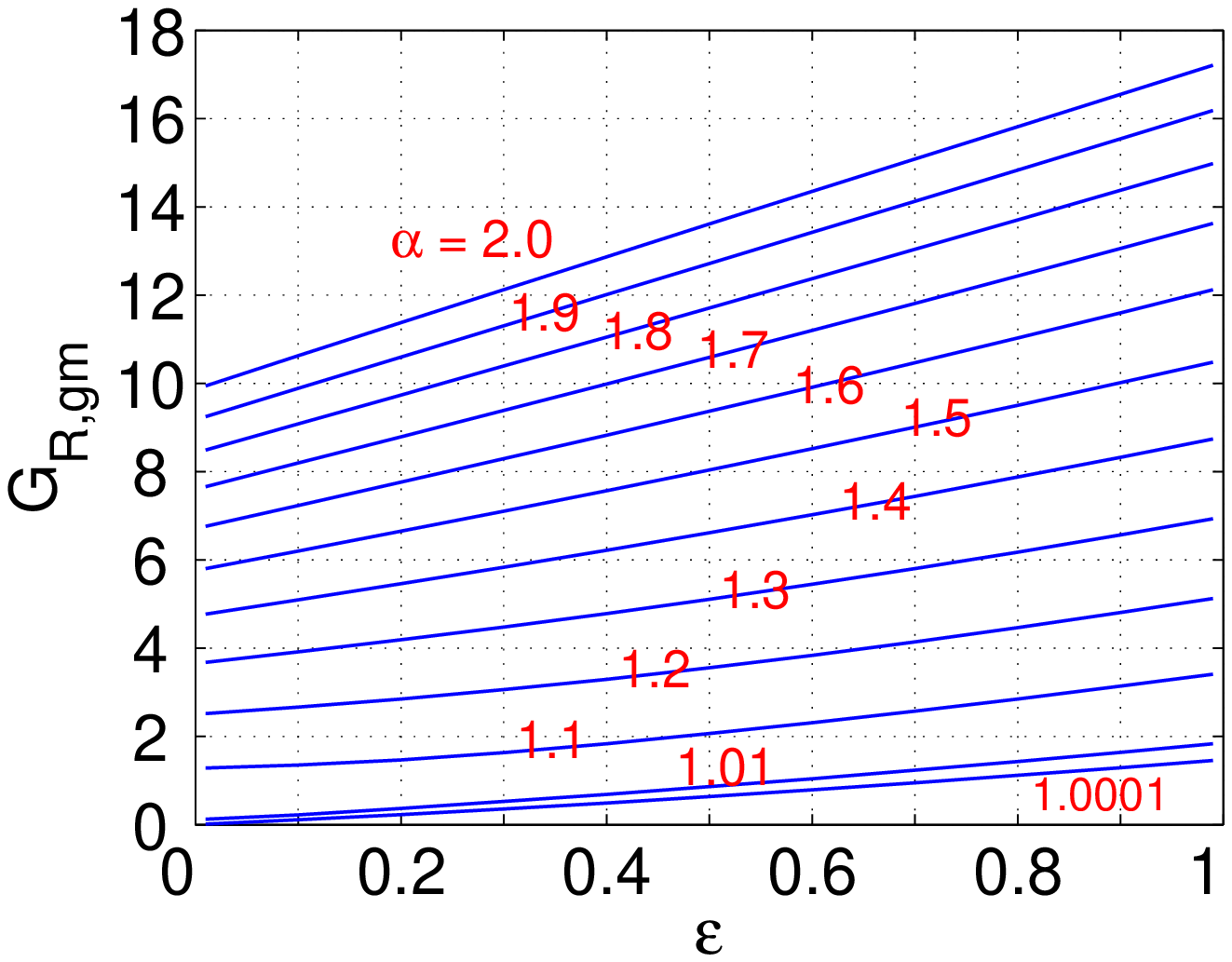}}}\vspace{-0.1in}
\mbox{
\subfigure[Left  bound, $\alpha<1$]{\includegraphics[width=1.6in]{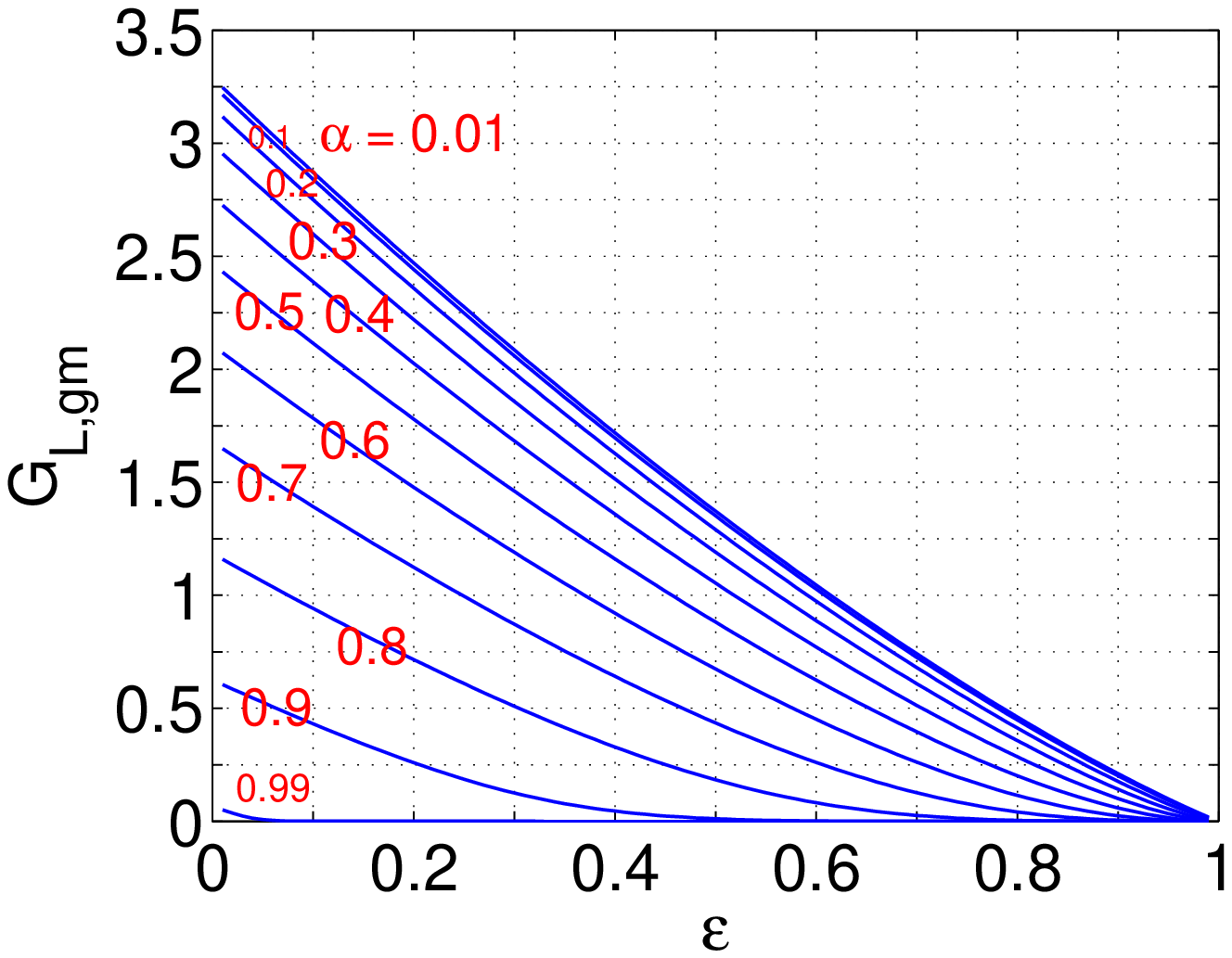}}\hspace{-0.1in}
\subfigure[Left  bound, $\alpha>1$]{\includegraphics[width=1.6in]{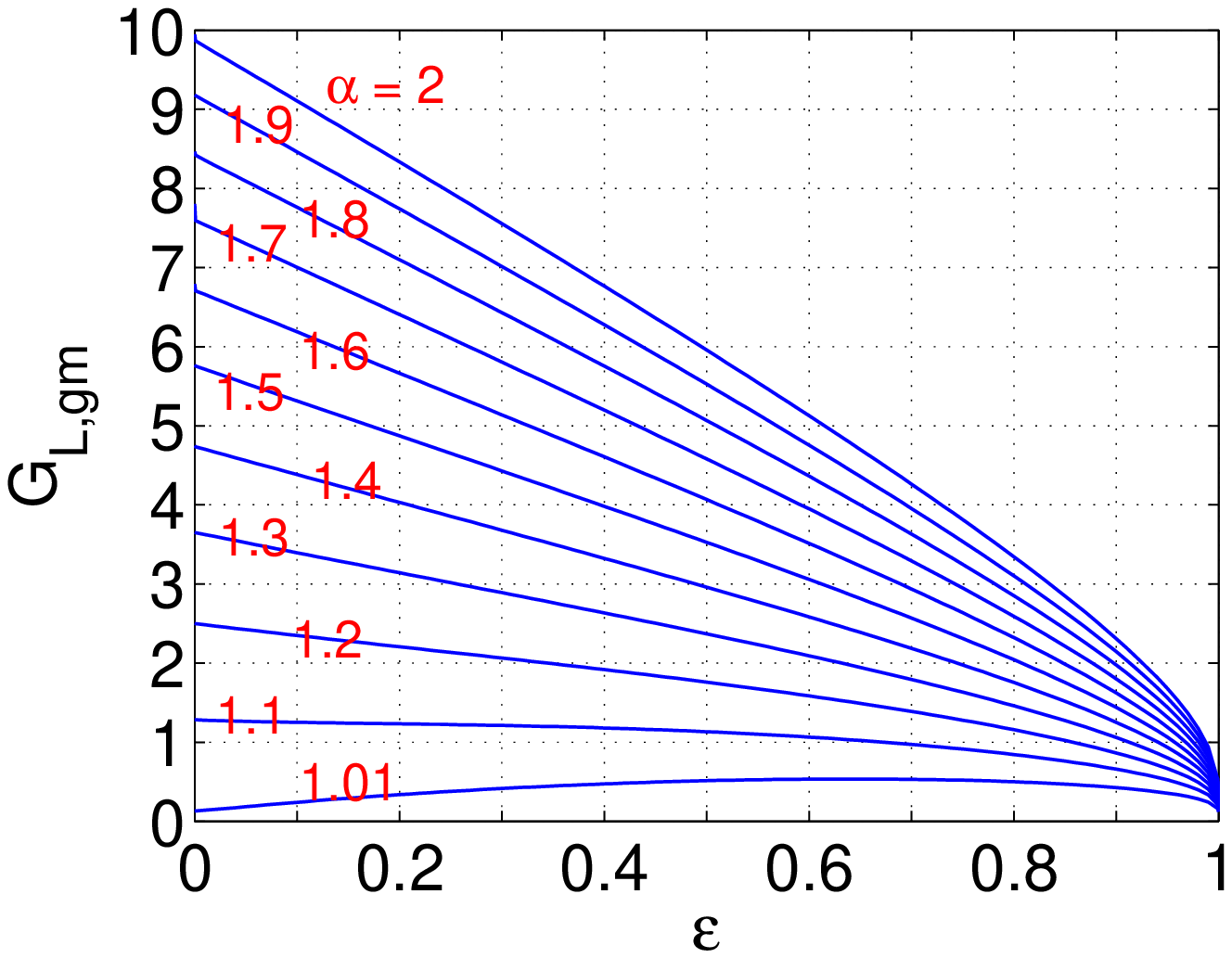}}}
\end{center}
\vspace{-0.3in}
\caption{The tail bound constants of $\hat{F}_{(\alpha),gm,b}$ in Lemma \ref{lem_gm_bounds}.
}\label{fig_G_gm}\vspace{-0.in}
\end{figure}

Lemma \ref{lem_gm_bounds}  provides the tail bounds for $\hat{F}_{(\alpha),gm,b}$ and Figure \ref{fig_G_gm} plots the tail bound constants. One can infer the tail bounds for $\hat{F}_{(\alpha),gm}$ from the monotonicity result (\ref{eqn_gm_asymp}).

\begin{lemma}\label{lem_gm_bounds}
The right tail bound:\vspace{-0.1in}
{\small\begin{align}\notag
&\mathbf{Pr}\left(\hat{F}_{(\alpha),gm,b} - F_{(\alpha)} \geq \epsilon F_{(\alpha)} \right) \leq
\exp\left(-k\frac{\epsilon^2}{G_{R,gm}}\right),  \ \
\epsilon>0,
\end{align}} and the left tail bound:\vspace{-0.1in}
{\small\begin{align}\notag
&\mathbf{Pr}\left(\hat{F}_{(\alpha),gm,b} - F_{(\alpha)} \leq -\epsilon F_{(\alpha)}\right)\leq \exp\left(-k \frac{\epsilon^2}{G_{L,gm}}\right), \ 0<\epsilon<1,
\end{align}}\vspace{-0.1in}
{\small\begin{align}\notag
&\frac{\epsilon^2}{G_{R,gm}} =
C_R \log(1+\epsilon) - C_R \gamma_e(\alpha-1)
 \\\notag
 & - \log\left(\cos\left(\frac{\kappa(\alpha)\pi C_R}{2}\right)
\frac{2}{\pi}\Gamma\left(\alpha C_R\right)\Gamma\left(1-C_R\right)\sin\left(\frac{\pi\alpha C_R}{2}\right)\right),\\\notag
&\frac{\epsilon^2}{G_{L,gm}}
=
-C_L \log(1-\epsilon)  + C_L\gamma_e(\alpha-1)+\log\alpha\\\notag
&\hspace{0.in} -  \log\left(\cos\left(\frac{\kappa(\alpha)\pi}{2}C_L\right)\Gamma\left(C_L\right)\right)+\log\left(   \Gamma\left(\alpha C_L\right)\cos\left(\frac{\pi\alpha C_L}{2}\right)\right).
\end{align}}\vspace{-0.05in}

\noindent $C_R$  and $C_L$ are solutions to\vspace{-0.1in}
{\small\begin{align}\notag\vspace{-0.in}
& -\gamma_e (\alpha -1)+\log(1+\epsilon)+ \frac{\kappa(\alpha)\pi}{2}{\tan\left(\frac{\kappa(\alpha)\pi}{2}C_R\right)}\\\notag
 &\hspace{0.5in}- \frac{\alpha\pi/2}{\tan\left(\frac{\alpha\pi }{2}C_R\right)}-
\psi\left(\alpha C_R\right)\alpha + \psi\left(1-C_R\right)=0,\\\notag
&\log(1-\epsilon) - \gamma_e(\alpha-1) - \frac{\kappa(\alpha)\pi}{2} \tan\left(\frac{\kappa(\alpha)\pi}{2}C_L\right)\\\notag
&\hspace{0.5in} +\frac{\alpha\pi}{2} { \tan\left(\frac{\alpha\pi}{2}C_L\right)}
 -\psi\left(\alpha C_L\right)\alpha+ \psi\left(C_L\right)=0.
\end{align}}
Here $\psi(z) = \frac{\Gamma^\prime(z)}{\Gamma(z)}$ is the ``Psi'' function.

\noindent\textbf{Proof:}\hspace{0.2in} See Appendix \ref{proof_lem_gm_bounds}. $\Box$
\end{lemma}

\vspace{-0.1in}

It is important to understand the behavior of the tail bounds as $\alpha = 1\pm\Delta \rightarrow 1$. ($\alpha=1-\Delta$ if $\alpha<1$; and $\alpha=1+\Delta$ if $\alpha>1$.) See more comments in Appendix \ref{app_moments}. Lemma \ref{lem_G_gm_rate} describes the precise rates of convergence.

\begin{lemma}\label{lem_G_gm_rate}
For fixed $\epsilon$, as $\alpha \rightarrow 1$ (i.e., $\Delta \rightarrow 0$),
{\small\begin{align}\notag\vspace{-0.1in}
&G_{R,gm}= \frac{\epsilon^2}{\log(1+\epsilon) - 2\sqrt{\Delta\log\left(1+\epsilon\right)}+o\left(\sqrt{\Delta}\right)},\\\notag
&\text{If}\  \alpha>1, \  \text{then}\\\notag
&G_{L,gm} = \frac{\epsilon^2}{-\log(1-\epsilon) - 2\sqrt{-2\Delta\log(1-\epsilon)} + o\left(\sqrt{\Delta}\right)},\\\notag
&\text{If}  \ \alpha<1, \ \text{then}\\\notag
&G_{L,gm} = \frac{\epsilon^2}{ \Delta\left(\exp\left(\frac{-\log(1-\epsilon)}{\Delta} -1 - \gamma_e\right)\right)+o\left(\Delta\exp\left(\frac{1}{\Delta}\right)\right)}.
\end{align}}
\noindent\textbf{Proof:} \ \  See Appendix \ref{proof_lem_G_gm_rate}. $\Box$
\end{lemma}

\begin{figure}[h]\vspace{-0.in}
\begin{center}
\mbox{
\subfigure[Right bound, $\alpha<1$]{\includegraphics[width=1.6in]{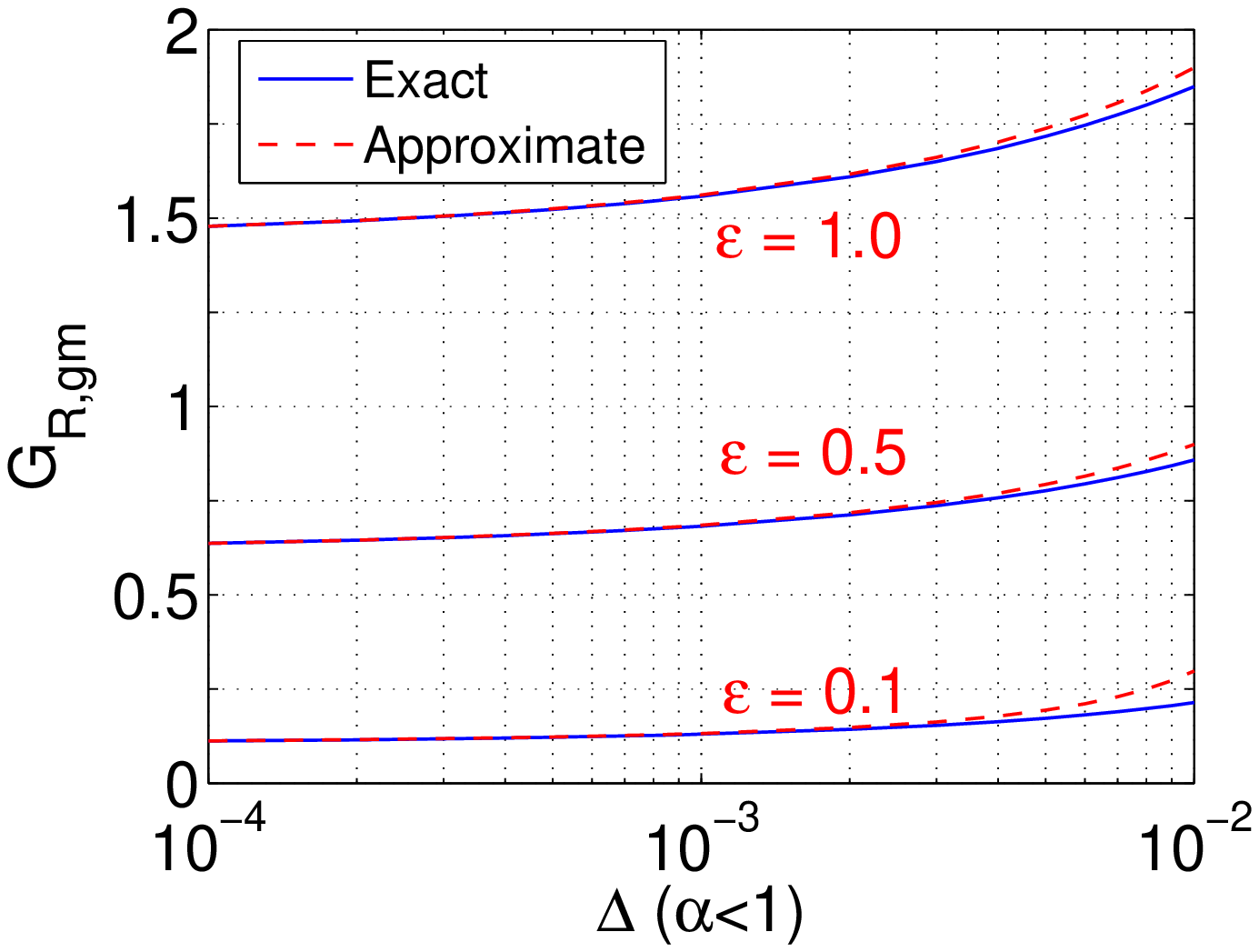}}\hspace{-0.1in}
\subfigure[Right bound, $\alpha>1$]{\includegraphics[width=1.6in]{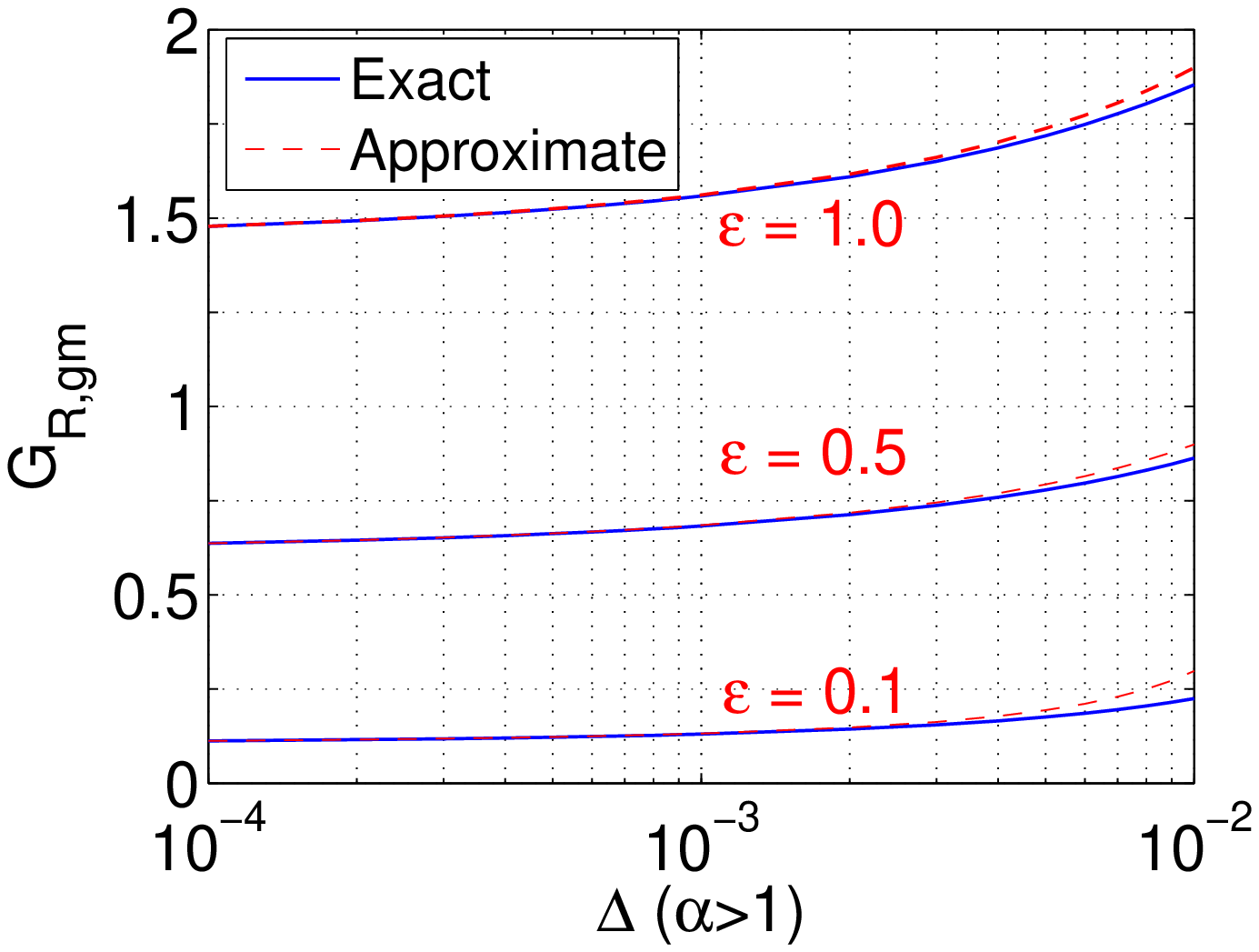}}}\vspace{-0.1in}
\mbox{
\subfigure[Left bound, $\alpha<1$]{\includegraphics[width=1.6in]{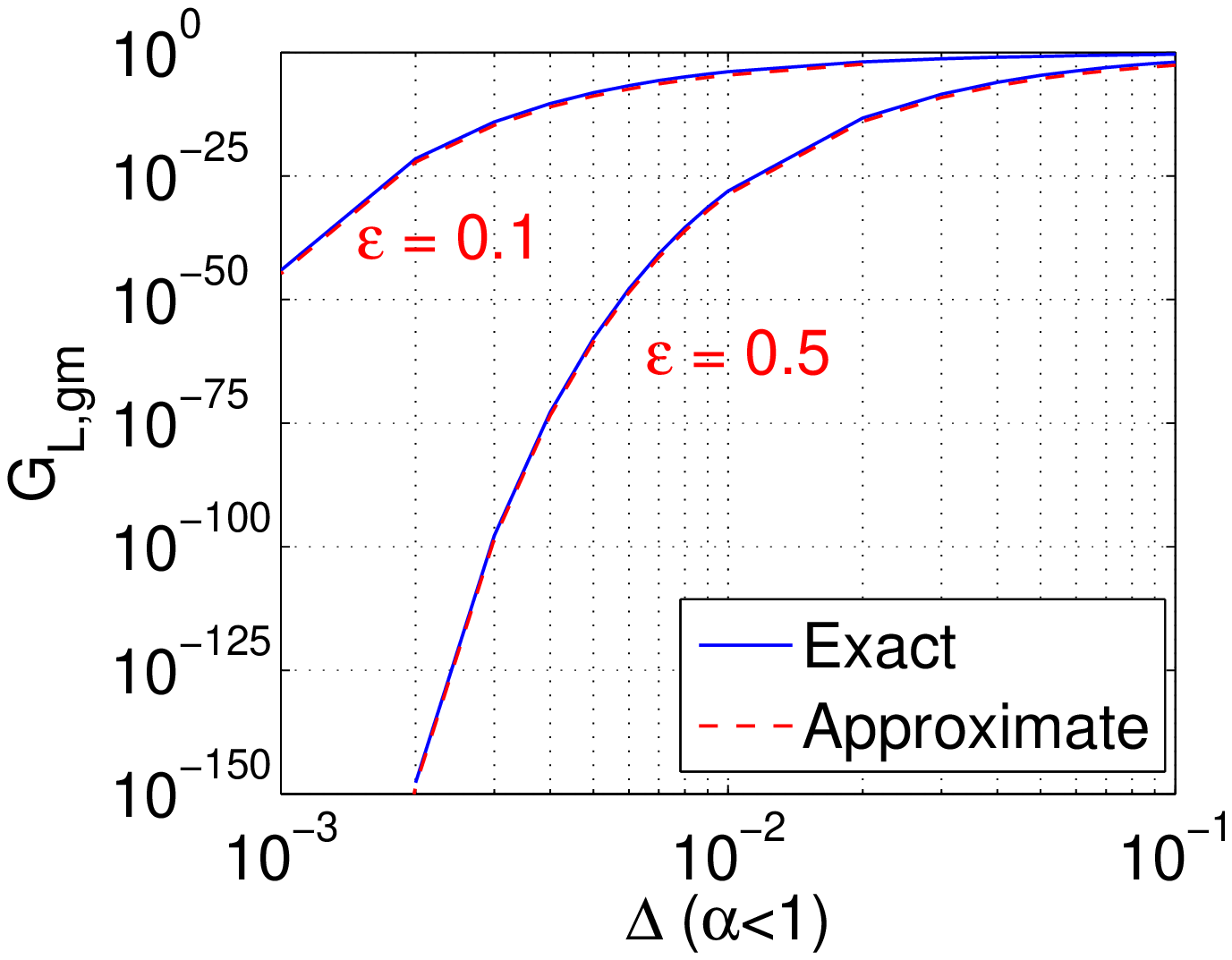}}\hspace{-0.1in}
\subfigure[Left bound, $\alpha>1$]{\includegraphics[width=1.6in]{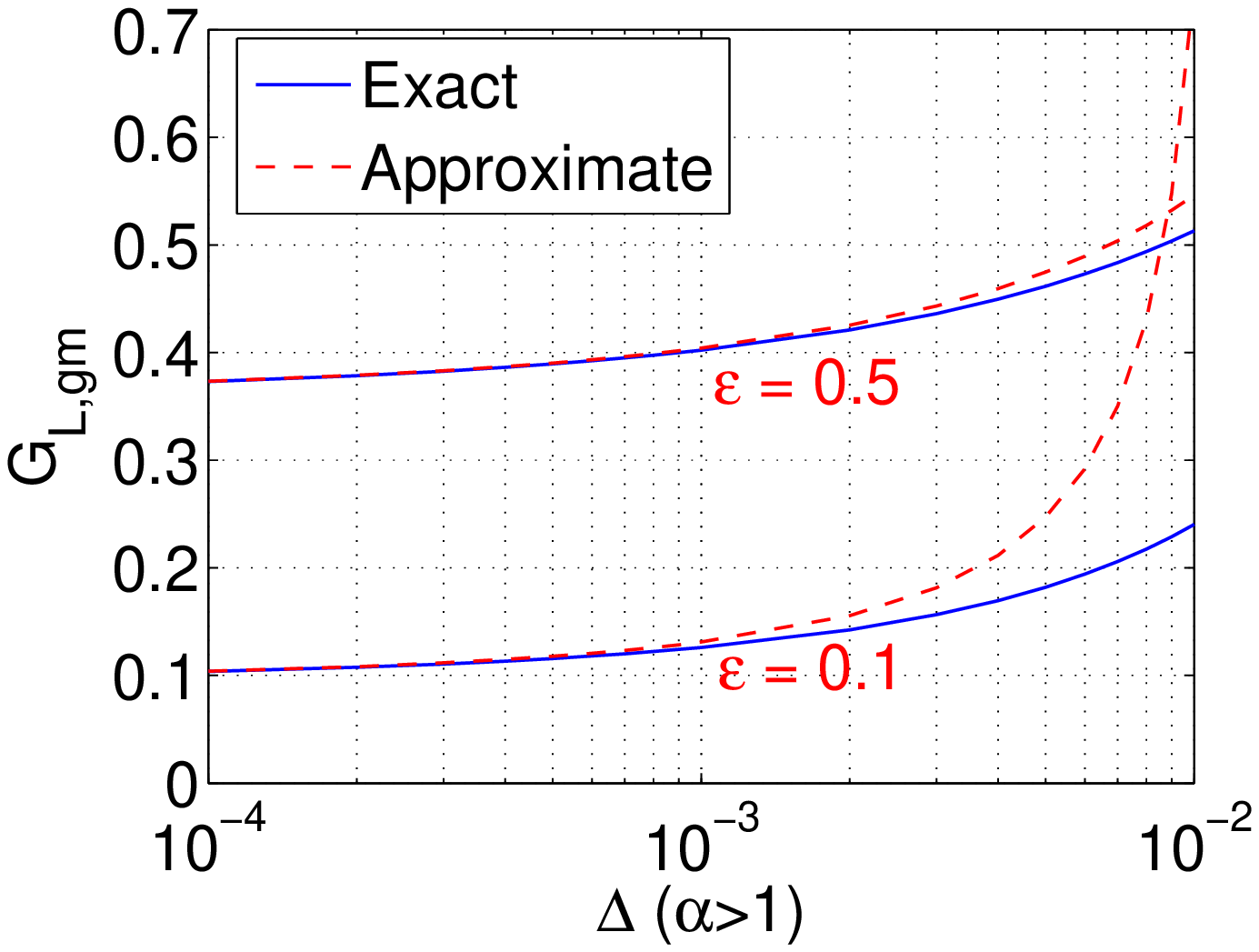}}}
\end{center}
\vspace{-0.3in}
\caption{The tail bound constants proved in Lemma \ref{lem_gm_bounds} and the approximations in Lemma \ref{lem_G_gm_rate}, for small $\Delta$.
}\label{fig_G_gm_approx}
\end{figure}

Figure \ref{fig_G_gm_approx} plots the constants for small values of $\Delta$, along with the approximations suggested in Lemma \ref{lem_G_gm_rate}. Since we usually consider $\epsilon$ should not be too large, we can write, as $\alpha\rightarrow 1$,  $G_{R,gm} = O\left(\epsilon\right)$ and $G_{L,gm} = O\left(\epsilon\right)$ if $\alpha>1$; both at the rate {\small$O\left(\sqrt{\Delta}\right)$}.  However, if $\alpha<1$, {\small$G_{L,gm} = O\left(\epsilon \exp\left(-\frac{\epsilon}{\Delta}\right)\right)$}, which is extremely fast.\\\vspace{-0.05in}

The sample complexity bound is then straightforward.\vspace{-0.04in}
\begin{lemma}\label{lem_JL}
Using the geometric mean estimator, it suffices to let $k = G\frac{1}{\epsilon^2}\log\left(\frac{2}{\delta}\right)$ so that the error will be within a $1\pm\epsilon$ factor with probability $1-\delta$, where $G = \max(G_{R,gm}, G_{L,gm})$. In the neighborhood of $\alpha = 1$, $k = O\left(\frac{1}{\epsilon}\log \frac{2}{\delta}\right)$ only.
\end{lemma}

\vspace{-0.06in}
\section{The Harmonic Mean Estimator}\label{sec_hm}
\vspace{-0.04in}

For $\alpha <1$, the {\em harmonic mean} estimator can considerably improve $\hat{F}_{(\alpha),gm}$. Unlike the {\em harmonic mean} estimator in \cite{Proc:Li_SODA08}, which is useful only for small $\alpha$ and has no exponential tail bounds except for $\alpha=0+$, the {\em harmonic mean} estimator in this study has very nice tail properties for all $0<\alpha<1$.

The {\em harmonic mean} estimator takes advantage of the fact that if $Z\sim S(\alpha<1, \beta =1, F_{(\alpha)})$, then $\text{E}\left(|Z|^\lambda\right)=\text{E}\left(Z^\lambda\right)$ exists for all $-\infty<\lambda <\alpha$.

\begin{lemma}\label{lem_hm}
Assume $k$ i.i.d. samples $x_j \sim S(\alpha<1, \beta=1, F_{(\alpha)})$, define the
harmonic mean estimator $\hat{F}_{(\alpha),hm}$,
{\small\begin{align}\notag\vspace{-0.05in}
\hat{F}_{(\alpha),hm} = \frac{k\frac{\cos\left(\frac{\alpha\pi}{2}\right)}{\Gamma(1+\alpha)}}{\sum_{j=1}^k|x_j|^{-\alpha}},
\end{align}}
and the bias-corrected harmonic mean estimator $\hat{F}_{(\alpha),hm,c}$,
{\small\begin{align}\notag
\hat{F}_{(\alpha),hm,c} = \frac{k\frac{\cos\left(\frac{\alpha\pi}{2}\right)}{\Gamma(1+\alpha)}}{\sum_{j=1}^k|x_j|^{-\alpha}}
\left(1- \frac{1}{k}\left(\frac{2\Gamma^2(1+\alpha)}{\Gamma(1+2\alpha)}-1\right) \right).
\end{align}}
The bias and variance of $\hat{F}_{(\alpha),hm,c}$ are
{\small\begin{align}\notag
&\text{E}\left(\hat{F}_{(\alpha),hm,c}\right) = F_{(\alpha)}+O\left(\frac{1}{k^2}\right),\\\notag
&\text{Var}\left(\hat{F}_{(\alpha),hm,c}\right) = \frac{F^{2}_{(\alpha)}}{k}\left(\frac{2\Gamma^2(1+\alpha)}{\Gamma(1+2\alpha)}-1\right) + O\left(\frac{1}{k^2}\right).
\end{align}}

The right tail bound of $\hat{F}_{(\alpha),hm}$ is, for $\epsilon>0$,
{\small\begin{align}\notag
&\mathbf{Pr}\left(
\hat{F}_{(\alpha),hm} -  F_{(\alpha)} \geq \epsilon F_{(\alpha)}\right)
\leq\exp\left(-k\left(\frac{\epsilon^2}{G_{R,hm}}  \right)\right), \\\notag
&\frac{\epsilon^2}{G_{R,hm}} = -\log   \left(\sum_{m=0}^\infty \frac{\Gamma^m(1+\alpha)}{\Gamma(1+m\alpha)}(-t_1^*)^m\right)
-\frac{t_1^*}{1+\epsilon},
\end{align}}
\noindent where $t_1^*$ is the solution to
{\small\begin{align}\notag
\frac{\sum_{m=1}^\infty(-1)^m m (t_1^*)^{m-1}\frac{\Gamma^m(1+\alpha)}{\Gamma(1+m\alpha)} }{\sum_{m=0}^\infty(-1)^m (t_1^*)^{m}\frac{\Gamma^m(1+\alpha)}{\Gamma(1+m\alpha)} } + \frac{1}{1+\epsilon} = 0.
\end{align}}

The left tail bound of $\hat{F}_{(\alpha),hm}$ is, for $0<\epsilon<1$,
{\small\begin{align}\notag
&\mathbf{Pr}\left(
\hat{F}_{(\alpha),hm} -  F_{(\alpha)} \leq -\epsilon F_{(\alpha)}\right)
\leq\exp\left(-k\left(\frac{\epsilon^2}{G_{L,hm}}\right)\right),\\\notag
&\frac{\epsilon^2}{G_{L,hm}} = -\log   \left(\sum_{m=0}^\infty \frac{\Gamma^m(1+\alpha)}{\Gamma(1+m\alpha)}(t_2^*)^m\right)
+\frac{t_2^*}{1-\epsilon}
\end{align}}
\noindent where $t_2^*$ is the solution to
{\small\begin{align}\notag
-\frac{\sum_{m=1}^\infty m (t_2^*)^{m-1}\frac{\Gamma^m(1+\alpha)}{\Gamma(1+m\alpha)} }{\sum_{m=0}^\infty (t_2^*)^{m}\frac{\Gamma^m(1+\alpha)}{\Gamma(1+m\alpha)} } + \frac{1}{1-\epsilon} = 0
\end{align}}

\noindent\textbf{Proof:}\hspace{0.2in} See Appendix \ref{proof_lem_hm}.   $\Box$.
\end{lemma}
\vspace{-0.25in}
\begin{figure}[h]
\begin{center}
\mbox{\subfigure[Right tail bound constant]{\includegraphics[width=1.6in]{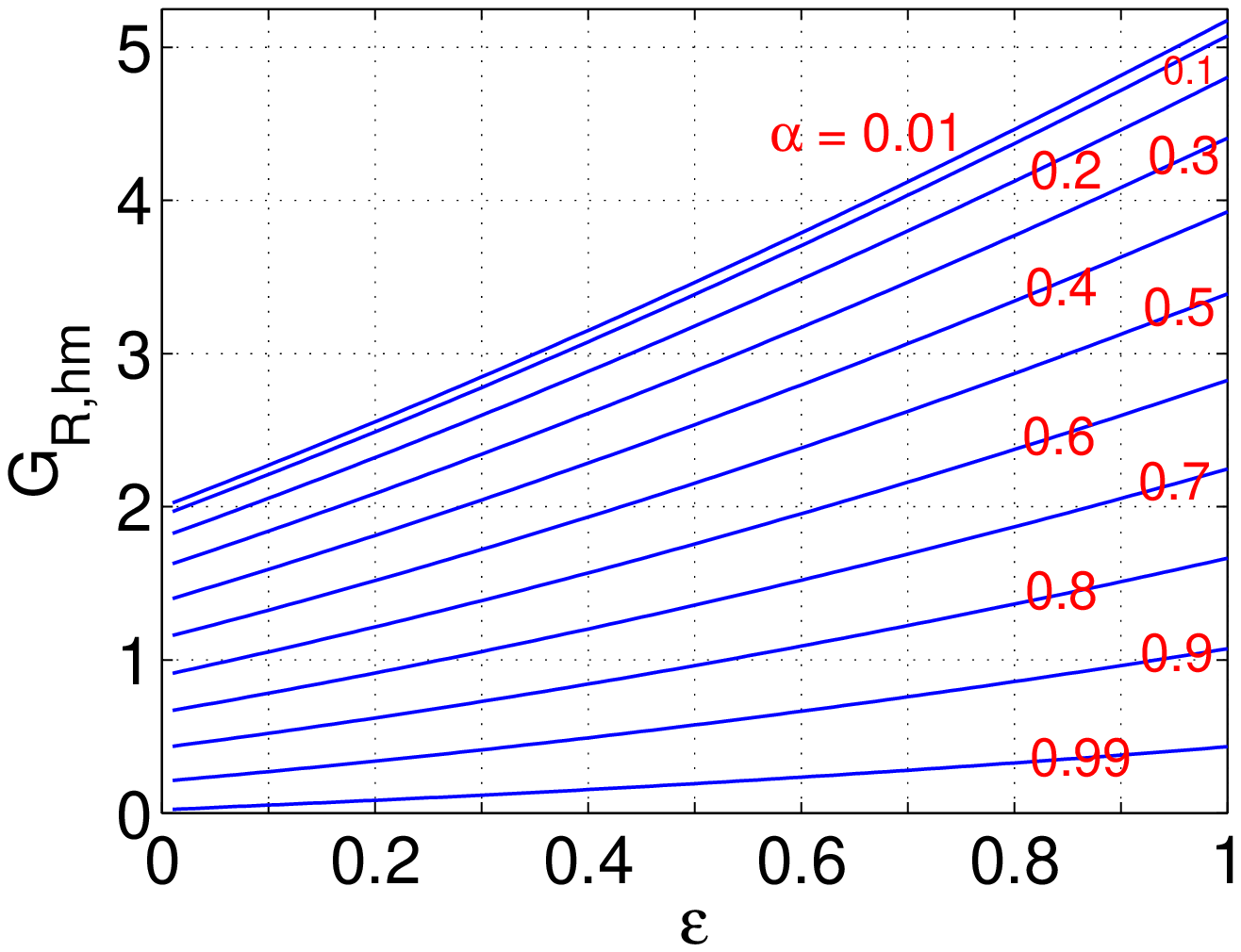}}\hspace{-0.1in}
\subfigure[Left tail bound constant]{\includegraphics[width=1.6in]{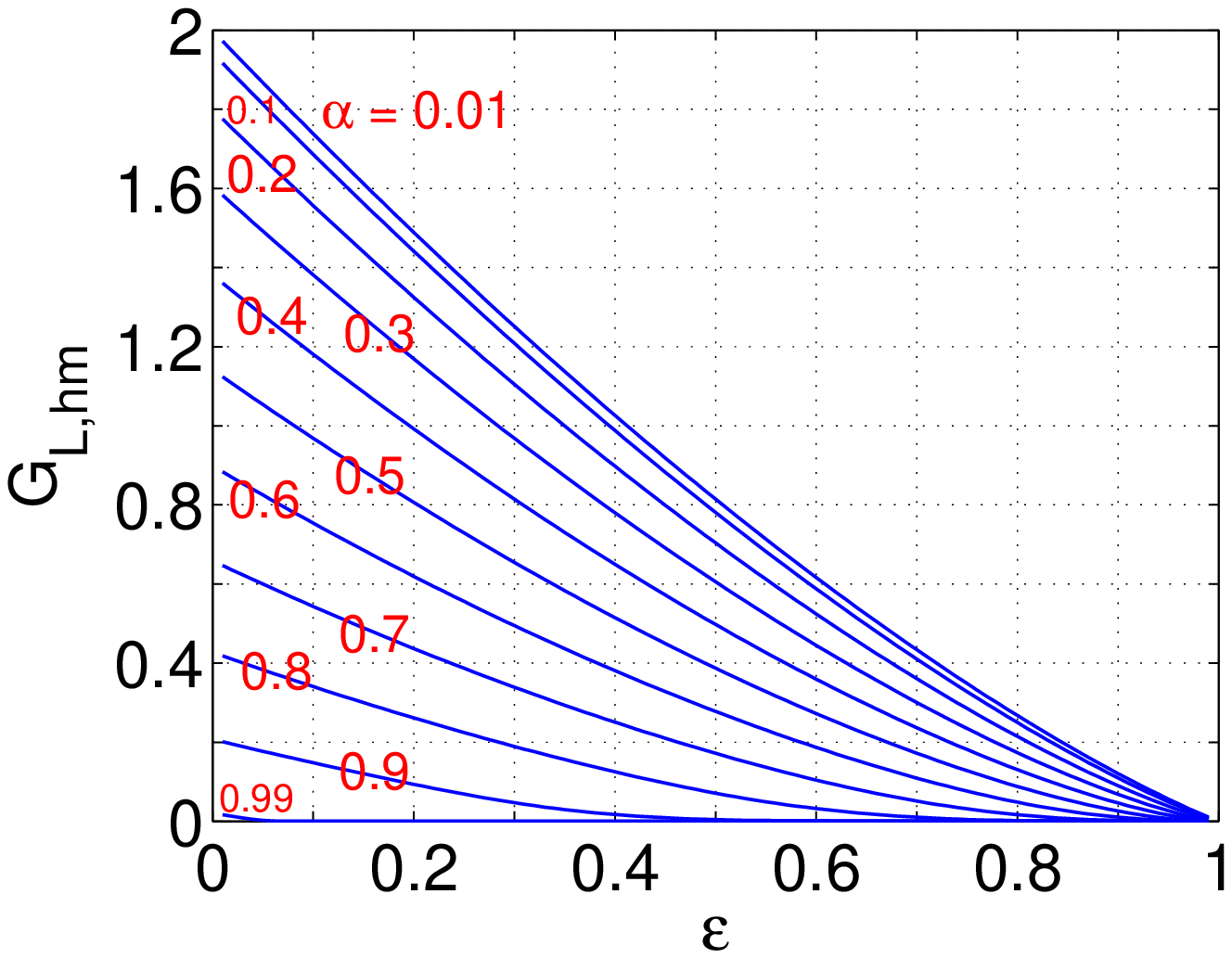}}}
\end{center}
\vspace{-0.3in}
\caption{The tail bound constants of $\hat{F}_{(\alpha),hm}$ in Lemma \ref{lem_hm}, which are considerably smaller, compared to Figure \ref{fig_G_gm}(a)(c). }\label{fig_G_hm}
\end{figure}

\vspace{-0.15in}
\section{The Logarithmic Norm and Distance}\label{sec_log}

The logarithmic norm and distance can  be important  in practice. Consider estimating the parameters from $D$ i.i.d. samples $x_i \sim Gamma(\theta,\gamma)$.
The density function is $f_X(x) = x^{\theta-1}\frac{\exp\left(-x/\gamma\right)}{\gamma^\theta\Gamma(\theta)}$, and the likelihood equation is
{\small\begin{align}\notag\vspace{-0.1in}
(\theta - 1)\sum_{i=1}^D\log x_i - \sum_{i=1}^D{x_i}/\gamma - D\theta\log(\gamma) - D\log\Gamma(\theta).
\end{align}}
If instead, $x_i \sim Pareto(\theta)$, $i = 1$ to $D$, then the density is $f_X(x) = \frac{\theta}{x^{\theta+1}}$, $x\geq 1$, and the likelihood equation is
{\small\begin{align}\notag\vspace{-0.1in}
D\log\theta - (\theta+1)\sum_{i=1}^D\log x_i.
\end{align}}\vspace{-0.15in}

Therefore, the logarithmic norm occurs at least in the content of maximum likelihood estimations of common distributions. Now, consider the data $x_i$'s are actually the elements of data streams $A_t[i]$'s. Estimating $\sum_{i=1}^D\log A_t[i]$ becomes an interesting and practically meaningful problem.

Our solution is based on the fact that, as $\alpha\rightarrow 0+$,
{\small\begin{align}\notag
\frac{D}{\alpha}\log\left(\frac{1}{D}\sum_{i=1}^DA_t[i]^\alpha\right)\rightarrow \sum_{i=1}^D\log A_t[i],
\end{align}}
which can be shown by L'H\'opital's rule. More precisely,
{\small\begin{align}\notag
&\left|\frac{D}{\alpha}\log\left(\frac{1}{D}\sum_{i=1}^DA_t[i]^\alpha\right) - \sum_{i=1}^D\log A_t[i]\right| \\\notag =&O\left(\frac{\alpha}{D}\left(\sum_{i=1}^D\log A_t[i]\right)^2\right) + O\left(\alpha\sum_{i=1}^D\log^2A_t[i]\right),
\end{align}}
which can be shown by Taylor expansions.

Therefore, we obtain one solution to approximating the logarithmic norm using very small $\alpha$.
Of course, we have assumed that $A_t[i]>0$ strictly. In fact, this also suggests an approach for approximating the logarithmic distance between two streams $\sum_{i=1}^D\log|A_t[i] - B_t[i]|$, provided  we use {\em symmetric stable random projections}.

The logarithmic distance can be  useful in machine learning practice with massive heavy-tailed data (either static or dynamic) such as image and text data. For those data, the usual $l_2$ distance would not be useful without ``term-weighting'' the data; and taking logarithm is one simple weighting scheme. Thus, our method provides a direct way to compute pairwise distances, taking into account data weighting automatically.

One may be also interested in the tail bounds, which, however, can not be expressed in terms of the logarithmic norm (or distance). Nevertheless, we can obtain, e.g.,
{\small \begin{align}\notag
&\mathbf{Pr}\left(\left[\frac{D}{\alpha}\log\left(\frac{1}{D}\hat{F}_{(\alpha),hm}\right)\right]\geq (1+\epsilon) \left[\frac{D}{\alpha}\log\left(\frac{1}{D}F_{(\alpha)}\right)\right]\right)\\\notag
\leq& \exp\left(-k\frac{\left(\left(F_{(\alpha)}/D\right)^{\epsilon}-1\right)^2}{G_{R,hm}}\right), \hspace{0.5in} \epsilon>0,\\\notag
&\mathbf{Pr}\left(\left[\frac{D}{\alpha}\log\left(\frac{1}{D}\hat{F}_{(\alpha),hm}\right)\right]\leq (1-\epsilon) \left[\frac{D}{\alpha}\log\left(\frac{1}{D}F_{(\alpha)}\right)\right]\right)\\\notag
\leq& \exp\left(-k\frac{\left(1-\left(D/F_{(\alpha)}\right)^{\epsilon}\right)^2}{G_{L,hm}}\right), \hspace{0.5in} 0<\epsilon<1
\end{align}}
If $\hat{F}_{(\alpha),gm}$ is used, we just replace the corresponding constants in the above expressions. If we are interested in the logarithmic distance, we simply apply {\em symmetric stable random projections} and use an appropriate estimator of the distance; the corresponding tail bounds will have same format.

\section{Conclusion}

Counting is a fundamental operation. In data streams $A_t[i]$, $i\in[1,D]$, counting the $\alpha$th frequency moments $F_{(\alpha)} = \sum_{i=1}^DA_t[i]^\alpha$ has been extensively studied. Our proposed {\em Compressed Counting (CC)} takes advantage of the fact that most data streams encountered in practice are non-negative, although they are subject to deletion and insertion. In fact, CC only requires that at the time $t$ for the evaluation, $A_t[i]\geq 0$; at other times, the data streams can actually go below zero.

{\em Compressed Counting} successfully captures the intuition that, when $\alpha = 1$,  a simple counter suffices, and when $\alpha=1\pm\Delta$ with small $\Delta$, an intelligent counting system should require  low space (continuously as a function of $\Delta$). The case with small $\Delta$ can be practically important. For example, $\Delta$ may be the ``decay rate'' or ``interest rate,'' which is usually small. CC can also be very useful for  statistical parameter estimation based on the {\em method of moments}. Also,  one can  approximate the entropy moment using the $\alpha$th moments with $\alpha = 1\pm\Delta$ and very small $\Delta$.

Compared with previous studies, e.g., \cite{Article:Indyk_JACM06,Proc:Li_SODA08}, {\em Compressed Counting} achieves, in a sense, an ``infinite improvement'' in terms of the asymptotic variances when $\Delta\rightarrow 0$. Two estimators based on the geometric mean and the harmonic mean are provided in this study, including their variances, tail bounds, and sample complexity bounds.

We analyze our sample complexity bound $k = G\frac{1}{\epsilon^2}\log\frac{2}{\delta}$ at the neighborhood of $\alpha = 1$ and  show  $G = O\left(\epsilon\right)$ at small $\Delta$. This implies that our bound at small $\Delta$ is actually $k =O\left(1/\epsilon\right)$ instead of $O\left(1/\epsilon^2\right)$, which is required in the
 Johnson-Lindenstrauss Lemma and its various analogs.

Finally, we propose a  scheme for approximating the logarithmic norm and the logarithmic distance, useful in statistical parameter estimation and machine learning practice.\\

We expect that new algorithms will soon be developed to take advantage of {\em Compressed Counting}. For example,
via private communications, we have learned that a group is vigorously developing algorithms using projections with {\small$\alpha=1\pm\Delta$} very close to 1, where {\small$\Delta$} is their important parameter.


\appendix
\vspace{-0.05in}
\section{An Example of Method of Moments}\label{app_moments}

We provide a (somewhat contrived) example of the {\em method of moments}. Suppose the observed data $x_i$'s are from data streams and suppose the data follows a gamma distribution $x_i \sim Gamma(\theta,1)$, i.i.d. Here, we only consider one parameter $\theta$ so that we can analyze the variance easily.

Suppose we estimate $\theta$ using the $\alpha$th moment. Because $\text{E}(x_i^\alpha) = \Gamma(\alpha+\theta)/\Gamma(\theta)$, we can solve for $\hat{\theta}$ from
{\scriptsize\begin{align}\notag
\frac{\Gamma(\alpha+\hat{\theta})}{\Gamma(\hat{\theta})} = \frac{1}{D}\sum_{i=1}^Dx_k^\alpha, \hspace{0.in} \Longrightarrow \text{Var}\left(\frac{\Gamma(\alpha+\hat{\theta})}{\Gamma(\hat{\theta})}\right) = \frac{1}{D}\left(\frac{\Gamma(2\alpha+\theta)}{\Gamma(\theta)} - \frac{\Gamma^2(\alpha+\theta)}{\Gamma^2(\theta)}\right)
\end{align}}
By the "delta method" (i.e., $\text{Var}(h(x))\approx \text{Var}(x)(h^\prime(\text{E}(x)))^2$) and using the implicit derivative of $\hat{\theta}$, we obtain
{\scriptsize\begin{align}\notag
\text{Var}\left(\hat{\theta}\right) \approx \frac{1}{D} \left(\frac{\Gamma(2\alpha+\theta)\Gamma(\theta)}{\Gamma^2(\alpha+\theta)} -1 \right)\frac{1}{\left(\psi(\alpha+\theta)-\psi(\theta)\right)^2}.
\end{align}}\vspace{-0.15in}

One can verify {\small$\text{Var}(\hat{\theta})$} increases monotonically with increasing {\small$\alpha\in[0,\infty)$}. Because $x_i$'s are from data streams,  we apply {\em Compressed Counting} for  the $\alpha$th moment. Suppose we consider the difference in  the estimation accuracy  at different  $\alpha$ is not important (because {\small$D$} is  large). Then we simply let $\alpha =1$. In case we need to estimate two parameters, we might choose {\small$\alpha=1$} and another $\alpha$  close to 1.

Now suppose we actually care about both the estimation accuracy (which favors smaller $\alpha$)  and the computational efficiency (which favors $\alpha=1$), we then need to balance this trade-off by choosing $\alpha$. To do so, we need to know the precise behavior of {\em Compressed Counting} in the neighborhood of $\alpha =1$, as well as the precise behavior of  $\hat{\theta}$, i.e., its tail bounds (not just variance). Thus, our analysis on the convergence rates in Lemma \ref{lem_G_gm_rate} will be very useful.

\vspace{-0.05in}
\section{Proof of Lemma \ref{lem_moments}}\label{proof_lem_moments}

Assume $Z \sim S(\alpha,\beta,F_{(\alpha)})$.  To prove $\textbf{E}\left(|Z|^\lambda\right)$ for $-1<\lambda <\alpha$, \cite[Theorem 2.6.3]{Book:Zolotarev_86} provided only a partial answer:
{\scriptsize\begin{align}\notag
&\int_0^\infty z^\lambda f_Z(z;\alpha,\beta_B,F_{(\alpha)}) dz \\\notag
=& F^{\lambda/\alpha}_{(\alpha)}\frac{\sin(\pi\rho\lambda)}{\sin(\pi\lambda)}\frac{\Gamma\left(1-\frac{\lambda}{\alpha}\right)}{\Gamma\left(1-\lambda\right)} \cos^{-\lambda/\alpha}\left(\pi\beta_B\kappa(\alpha)/2\right)
\end{align}}
where we denote
{\scriptsize\begin{align} \notag
\kappa(\alpha) = \alpha \ \ \  \text{if}  \ \ \ \alpha<1, \ \  \ \text{and} \  \ \kappa(\alpha)=2-\alpha\ \  \ \text{if} \ \ \alpha>1,
\end{align}}
and  according to the parametrization used in \cite[I.19, I.28]{Book:Zolotarev_86}:
{\scriptsize\begin{align}\notag
  \beta_B = \frac{2}{\pi\kappa(\alpha)}\tan^{-1}\left(\beta\tan\left(\frac{\pi\alpha}{2}\right)\right),\hspace{0.1in}
  \rho  = \frac{1-\beta_B\kappa(a)/\alpha}{2}.
\end{align}}
Note that
{\scriptsize\begin{align}\notag
&\cos^{-\lambda/\alpha}\left(\pi\beta_B\kappa(\alpha)/2\right) = \left(1 + \tan^2\left(\pi\beta_B\kappa(\alpha)/2\right)\right)^{\frac{\lambda}{2\alpha}}\\\notag
 =&  \left(1 + \tan^2\left(\tan^{-1}\left(\beta\tan\left(\frac{\pi\alpha}{2}\right)\right)\right)\right)^{\frac{\lambda}{2\alpha}}\\\notag
 =& \left(1 + \beta^2\tan^2\left(\frac{\pi\alpha}{2}\right)\right)^{\frac{\lambda}{2\alpha}}.
\end{align}}
Therefore, for $-1<\lambda<\alpha$,
{\scriptsize\begin{align}\notag
&\int_0^\infty z^\lambda f_Z(z;\alpha,\beta_B,F_{(\alpha)}) dz\\\notag
 =& F^{\lambda/\alpha}_{(\alpha)} \frac{\sin(\pi\rho\lambda)}{\sin(\pi\lambda)}\frac{\Gamma\left(1-\frac{\lambda}{\alpha}\right)}{\Gamma\left(1-\lambda\right)}
\left(1 + \beta^2\tan^2\left(\frac{\pi\alpha}{2}\right)\right)^{\frac{\lambda}{2\alpha}}.
\end{align}}

To compute $\text{E}\left(|Z|^\lambda\right)$, we take advantage of a useful property of the stable density function\cite[page 65]{Book:Zolotarev_86}:
{\scriptsize\begin{align}\notag
f_Z(- z;\alpha,\beta_B,F_{(\alpha)}) = f_Z(z;\alpha,-\beta_B,F_{(\alpha)}).
\end{align}
\begin{align}\notag
&\text{E}\left(|Z|^\lambda\right)\\\notag
 =& \int_{-\infty}^0 (-z)^\lambda f_Z(z;\alpha,\beta_B,F_{(\alpha)}) dz + \int_0^\infty z^\lambda f_Z(z;\alpha,\beta_B,F_{(\alpha)}) dz\\\notag
=& \int_0^{\infty} z^\lambda f_Z(z;\alpha,-\beta_B,F_{(\alpha)}) dz + \int_0^\infty z^\lambda f_Z(z;\alpha,\beta_B,F_{(\alpha)}) dz\\\notag
=&\frac{F^{\lambda/\alpha}_{(\alpha)}}{\sin(\pi\lambda)}\frac{\Gamma\left(1-\frac{\lambda}{\alpha}\right)}{\Gamma\left(1-\lambda\right)}
\left(1 + \beta^2\tan^2\left(\frac{\pi\alpha}{2}\right)\right)^{\frac{\lambda}{2\alpha}}\\\notag
&\times\left(
\sin\left(\pi\lambda\frac{1-\beta_B\kappa(\alpha)/\alpha}{2}\right)+
\sin\left(\pi\lambda\frac{1+\beta_B\kappa(\alpha)/\alpha}{2}\right)
\right)\\\notag
=&\frac{F^{\lambda/\alpha}_{(\alpha)}}{\sin(\pi\lambda)}\frac{\Gamma\left(1-\frac{\lambda}{\alpha}\right)}{\Gamma\left(1-\lambda\right)} \left(1 + \beta^2\tan^2\left(\frac{\pi\alpha}{2}\right)\right)^{\frac{\lambda}{2\alpha}}\\\notag
&\times\left(
2\sin\left(\frac{\pi\lambda}{2}\right)\cos\left(\frac{\pi\lambda}{2}\beta_B\kappa(\alpha)/\alpha\right)\right)\\\notag
=&\frac{F^{\lambda/\alpha}_{(\alpha)}}{\cos(\pi\lambda/2)}\frac{\Gamma\left(1-\frac{\lambda}{\alpha}\right)}{\Gamma\left(1-\lambda\right)} \left(1 + \beta^2\tan^2\left(\frac{\pi\alpha}{2}\right)\right)^{\frac{\lambda}{2\alpha}}\\\notag
&\times\cos\left(\frac{\lambda}{\alpha}\tan^{-1}\left(\beta\tan\left(\frac{\pi\alpha}{2}\right)\right)   \right) \\\notag
=&F^{\lambda/\alpha}_{(\alpha)}\left(1 + \beta^2\tan^2\left(\frac{\pi\alpha}{2}\right)\right)^{\frac{\lambda}{2\alpha}}
\cos\left(\frac{\lambda}{\alpha}\tan^{-1}\left(\beta\tan\left(\frac{\pi\alpha}{2}\right)\right)   \right)\\\notag
&\hspace{0.5in}\times\left(\frac{2}{\pi}\sin\left(\frac{\pi}{2}\lambda\right)\Gamma\left(1-\frac{\lambda}{\alpha}\right)\Gamma\left(\lambda\right)\right),
\end{align}}
which can be simplified when $\beta = 1$, to be
{\scriptsize\begin{align}\notag
\textbf{E}\left(|Z|^\lambda\right) = {F}_{(\alpha)}^{\lambda/\alpha} \frac{\cos\left(\frac{\kappa(\alpha)}{\alpha}\frac{\lambda\pi}{2}\right)}
{\cos^{\lambda/\alpha}\left(\frac{\kappa(\alpha)\pi}{2}\right)}
\left(\frac{2}{\pi}\sin\left(\frac{\pi}{2}\lambda\right)\Gamma\left(1-\frac{\lambda}{\alpha}\right)\Gamma\left(\lambda\right)\right).
\end{align}}

The final task is to  show that when $\alpha <1$ and $\beta=1$, $\text{E}\left(|Z|^\lambda\right)$ exists for all $-\infty<\lambda<\alpha$, not just $-1<\lambda<\alpha$. This is an extremely useful property.

Note that when $\alpha<1$ and $\beta = 1$, $Z$ is always non-negative. As shown in the proof of \cite[Theorem 2.6.3]{Book:Zolotarev_86},
{\scriptsize\begin{align}\notag
&\text{E}\left(|Z|^\lambda\right)
= F_{(\alpha)}^{\lambda/\alpha} \cos^{-\lambda/\alpha}\left(\frac{\pi\alpha}{2}\right) \frac{1}{\pi}
\text{Im}
\int_0^\infty z^\lambda \int_0^\infty \\\notag
&\exp\left( - zu \exp(\sqrt{-1}\pi/2) - u^\alpha \exp(-\sqrt{-1}\pi\alpha/2)+\frac{\sqrt{-1}\pi}{2}\right)dudz\\\notag
=&F_{(\alpha)}^{\lambda/\alpha} \cos^{-\lambda/\alpha}\left(\frac{\pi\alpha}{2}\right) \frac{1}{\pi}\times\\\notag \text{Im}
&\int_0^\infty \int_0^\infty z^\lambda \exp\left( - zu\sqrt{-1} - u^\alpha \exp(-\sqrt{-1}\pi\alpha/2)\right)\sqrt{-1}dudz.
\end{align}}
The only thing we need to check is that in the proof of \cite[Theorem 2.6.3]{Book:Zolotarev_86}, the condition for Fubini's theorem (to exchange order of integration) still holds when $-\infty<\alpha <1$, $\beta = 1$, and $\lambda<-1$. We can show
{\scriptsize \begin{align}\notag
&\int_0^\infty \int_0^\infty \left| z^\lambda \exp\left( - zu\sqrt{-1} - u^\alpha \exp(-\sqrt{-1}\pi\alpha/2)\right)\sqrt{-1}\right|dudz\\\notag
 =&\int_0^\infty\int_0^\infty z^\lambda \left|\exp\left(-u^\alpha\cos(\pi\alpha/2)+ \sqrt{-1}u^\alpha\sin(\pi\alpha/2)\right)\right|dudz\\\notag
 =&\int_0^\infty\int_0^\infty z^\lambda \exp\left(-u^\alpha\cos(\pi\alpha/2)\right)dudz<\infty,
 \end{align}}
 \noindent provided $\lambda<-1$ ($\lambda \neq -1, -2, -3, ....$) and $\cos(\pi\alpha/2)>0$, i.e., $\alpha<1$. Note that $|\exp(\sqrt{-1}x)|=1$ always and Euler's formula: $\exp(\sqrt{-1}x) = \cos(x) + \sqrt{-1}\sin(x)$ is frequently used to simplify the algebra.

Once we show that Fubini's condition is satisfied, we can exchange the order of integration and the rest  follows from the proof of
\cite[Theorem 2.6.3]{Book:Zolotarev_86}. Because of continuity, the ``singularity points'' $\lambda = -1, -2, -3,...$ do not matter.

\section{Proof of Lemma \ref{lem_gm_moments}}\label{proof_lem_gm_moments}

We first show that, for any fixed $t$, as $k\rightarrow\infty$,
{\scriptsize\begin{align}
&\text{E}\left(\left(
\hat{F}_{(\alpha),gm} \notag
\right)^t\right) \\\notag
=& F_{(\alpha)}^t\frac{
\cos^{k}\left(\frac{\kappa(\alpha)\pi}{2k}t\right) \left[\frac{2}{\pi}\sin\left(\frac{\pi\alpha}{2k}t\right)
\Gamma\left(1-\frac{t}{k}\right)\Gamma\left(\frac{\alpha}{k}t\right)\right]^{k}}
{
\cos^{kt}\left(\frac{\kappa(\alpha)\pi}{2k}\right) \left[\frac{2}{\pi}\sin\left(\frac{\pi\alpha}{2k}\right)
\Gamma\left(1-\frac{1}{k}\right)\Gamma\left(\frac{\alpha}{k}\right)\right]^{kt}
}\\\notag
=& F_{(\alpha)}^t\exp\left( \frac{1}{k}\frac{\pi^2(t^2-t)}{24}\left(\alpha^2+2-3\kappa^2(\alpha)\right)+O\left(\frac{1}{k^2}\right)\right).
\end{align}}

In \cite{Proc:Li_SODA08}, it was proved that, as $k\rightarrow\infty$,
{\scriptsize\begin{align}\notag
&\frac{\left[\frac{2}{\pi}\sin\left(\frac{\pi\alpha}{2k}t\right)
\Gamma\left(1-\frac{t}{k}\right)\Gamma\left(\frac{\alpha}{k}t\right)\right]^{k}}
{ \left[\frac{2}{\pi}\sin\left(\frac{\pi\alpha}{2k}\right)
\Gamma\left(1-\frac{1}{k}\right)\Gamma\left(\frac{\alpha}{k}\right)\right]^{kt}
}\\\notag
=&1+\frac{1}{k}\frac{\pi^2(t^2-t)}{24}\left(\alpha^2+2\right)+O\left(\frac{1}{k^2}\right)\\\notag
=&\exp\left( \frac{1}{k}\frac{\pi^2(t^2-t)}{24}\left(\alpha^2+2\right)+O\left(\frac{1}{k^2}\right)\right).
\end{align}}
Using the infinite product representation of cosine\cite[1.43.3]{Book:Gradshteyn_00}
{\scriptsize\begin{align}\notag
\cos(z) = \prod_{s=0}^\infty\left(1 - \frac{4z^2}{(2s+1)^2\pi^2}\right),
\end{align}}
\noindent we can rewrite
{\scriptsize\begin{align}\notag
&\frac{\cos^{k}\left(\frac{\kappa(\alpha)\pi}{2k}t\right) } { \cos^{kt}\left(\frac{\kappa(\alpha)\pi}{2k}\right)  }
 = \prod_{s=0}^\infty \left(
1 - \frac{\kappa^2(\alpha)t^2}{(2s+1)^2k^2}\right)^k\left(
1 - \frac{\kappa^2(\alpha)}{(2s+1)^2k^2}\right)^{-kt}\\\notag
=&\prod_{s=0}^\infty \left(
\left( 1 - \frac{\kappa^2(\alpha)t^2}{(2s+1)^2k^2}\right)
\left(
1 + t\frac{\kappa^2(\alpha)}{(2s+1)^2k^2}+O\left(\frac{1}{k^3}\right)\right)\right)^k\\\notag
=&\prod_{s=0}^\infty
\left( 1 - \frac{\kappa^2(\alpha)(t^2-t)}{(2s+1)^2k^2} +O\left(\frac{1}{k^3}\right)\right)^k
=\prod_{s=0}^\infty
\left( 1 - \frac{\kappa^2(\alpha)(t^2-t)}{(2s+1)^2k} +O\left(\frac{1}{k^2}\right)\right)\\\notag
=&\exp\left(\sum_{s=0}^\infty
\log \left( 1 - \frac{\kappa^2(\alpha)(t^2-t)}{(2s+1)^2k} +O\left(\frac{1}{k^2}\right)\right)
\right)\\\notag
=&\exp\left(-\frac{\kappa^2(\alpha)}{k}(t^2-t) \sum_{s=0}^\infty \frac{1}{(2s+1)^2} +O\left(\frac{1}{k^2}\right)\right)
\\\notag
=&\exp\left(-\frac{\kappa^2(\alpha)}{k}(t^2-t) \frac{\pi^2}{8}+O\left(\frac{1}{k^2}\right)\right),
\end{align}}
\noindent which, combined with the result in \cite{Proc:Li_SODA08}, yields the desired expression. \\

The next task is to show
{\scriptsize\begin{align}\notag
\left[\cos\left(\frac{\kappa(\alpha)\pi}{2k}\right)\frac{2}{\pi}\Gamma\left(\frac{\alpha}{k}\right)\Gamma\left(1-\frac{1}{k}\right)\sin\left(\frac{\pi}{2}\frac{\alpha}{k}\right)\right]^k
\rightarrow \exp\left(-\gamma_e\left(\alpha-1\right)\right),
\end{align}}
\noindent monotonically as $k\rightarrow\infty$, where {\small$\gamma_e = 0.577215665...$}, is Euler's constant.
In \cite{Proc:Li_SODA08}, it was proved that, as $k\rightarrow\infty$,
{\scriptsize\begin{align}\notag
\left[\frac{2}{\pi}\Gamma\left(\frac{\alpha}{k}\right)\Gamma\left(1-\frac{1}{k}\right)\sin\left(\frac{\pi}{2}\frac{\alpha}{k}\right)\right]^k
\rightarrow \exp\left(-\gamma_e\left(\alpha-1\right)\right),
\end{align}}
\noindent monotonically.  In this study, we need to consider instead
{\scriptsize\begin{align}\notag\label{eqn_proof_gm_coefficent}
&\left[\cos\left(\frac{\kappa(\alpha)\pi}{2k}\right)\frac{2}{\pi}\Gamma\left(\frac{\alpha}{k}\right)\Gamma\left(1-\frac{1}{k}\right)\sin\left(\frac{\pi}{2}\frac{\alpha}{k}\right)\right]^k
\\
=&\left[2\cos\left(\frac{\kappa(\alpha)\pi}{2k}\right)\frac{\Gamma\left(\frac{\alpha}{k}\right)\sin\left(\frac{\pi\alpha}{2k}\right)}{\Gamma\left(\frac{1}{k}\right)\sin\left(\frac{\pi}{k}\right)} \right]^k
\end{align}}
(Note Euler's reflection formula $\Gamma(z)\Gamma(1-z) = \frac{\pi}{\sin(\pi z)}$.)
The additional term {\scriptsize$\left[\cos\left(\frac{\kappa(\alpha)\pi}{2k}\right)\right]^k = 1 - O\left(\frac{1}{k}\right)$}. Therefore,
{\scriptsize\begin{align}\notag
&\left[\cos\left(\frac{\kappa(\alpha)\pi}{2k}\right)\frac{2}{\pi}\Gamma\left(\frac{\alpha}{k}\right)\Gamma\left(1-\frac{1}{k}\right)\sin\left(\frac{\pi}{2}\frac{\alpha}{k}\right)\right]^k
\rightarrow \exp\left(-\gamma_e\left(\alpha-1\right)\right).
\end{align}}

To show the monotonicity, however, we have to use some different techniques from \cite{Proc:Li_SODA08}. The reason is because the additional term
$\left[\cos\left(\frac{\kappa(\alpha)\pi}{2k}\right)\right]^k$ increases (instead of decreasing) monotonically with increasing $k$.

First, we consider $\alpha>1$, i.e., $\kappa(\alpha) = 2-\alpha<1$. For simplicity, we take  logarithm of (\ref{eqn_proof_gm_coefficent}) and replace $1/k$ by $t$, where $0\leq t\leq 1/2$ (recall $k\geq 2$). It suffices to show that $g(t)$ increases with increasing $t \in [0,1/2]$, where
{\scriptsize\begin{align}\notag
&g(t) =\frac{1}{t}W(t),\\\notag
&W(t) = \log\left(\cos\left(\frac{\kappa(\alpha)\pi}{2}t\right)\right) + \log\left(\Gamma\left(\alpha t\right)\right) + \log\left(\sin\left(\frac{\pi\alpha}{2}t\right)\right)\\\notag
 &\hspace{0.5in}- \log\left(\Gamma\left(t\right)\right) - \log\left(\sin\left(\pi t\right)\right) + \log(2).
\end{align}}
Because $g^\prime(t) = \frac{1}{t}W^\prime(t) - \frac{1}{t^2}W(t)$, to show $g^\prime(t) \geq 0$ in $t \in [0,1/2]$, it suffices to show
{\scriptsize\begin{align}\notag
tW^\prime(t) - W(t) \geq 0.
\end{align}}
One can check that $tW^\prime(t)\rightarrow 0$ and $W(t)\rightarrow0$, as $t\rightarrow0+$.
{\scriptsize\begin{align}\notag
W^\prime(t) =& -\tan\left(\frac{\kappa(\alpha)\pi}{2}t\right)\left(\frac{\kappa\pi}{2}\right) + \psi\left(\alpha t\right)\alpha + \frac{1}{\tan\left(\frac{\pi\alpha}{2}t\right)}\left(\frac{\alpha\pi}{2}\right)\\\notag
 &\hspace{0.5in}- \psi(t) - \frac{1}{\tan(\pi t)}\pi.
\end{align}}
\noindent Here $\psi(x) = \frac{\partial \log(\Gamma(x))}{\partial x}$ is the ``Psi'' function.
Therefore, to show $tW^\prime(t) - W(t) \geq 0$, it suffices to show that $tW^\prime(t) - W(t)$ is an increasing function of $t\in [0,1/2]$, i.e.,
{\scriptsize\begin{align}\notag
&\left(tW^\prime(t) - W(t)\right)^\prime = W^{\prime\prime}(t) \geq 0, \ \ \ \text{i.e.,}\\\notag
&W^{\prime\prime}(t) = -\sec^2\left(\frac{\kappa(\alpha)\pi}{2}t\right)\left(\frac{\kappa(\alpha)\pi}{2}\right)^2 + \psi^\prime(\alpha t)\alpha^2\\\notag
 &\hspace{0.5in}- \csc^2\left(\frac{\pi\alpha}{2}t\right)\left(\frac{\pi\alpha}{2}\right)^2 - \psi^\prime(t) +\csc^2(\pi t)\pi^2 \geq 0.
\end{align}}

Using series representation of $\psi(x)$ \cite[8.363.8]{Book:Gradshteyn_00}, we show
{\scriptsize\begin{align}\notag
\psi^\prime\left(\alpha t\right)\alpha^2 - \psi^{\prime}(t) =& \sum_{s=0}^\infty \frac{\alpha^2}{(\alpha t + s)^2} - \sum_{s=0}^\infty \frac{1}{(t + s)^2}\\\notag =& \sum_{s=0}^\infty \left(\frac{1}{(t + s/\alpha)^2} - \frac{1}{(t + s)^2}\right) \geq 0,
\end{align}}
\noindent because we consider $\alpha >1$. Thus, it suffices to show that
{\scriptsize\begin{align}\notag
& Q(t;\alpha) = -\sec^2\left(\frac{\kappa(\alpha)\pi}{2}t\right)\left(\frac{\kappa(\alpha)\pi}{2}\right)^2\\\notag
  &\hspace{0.5in}- \csc^2\left(\frac{\pi\alpha}{2}t\right)\left(\frac{\pi\alpha}{2}\right)^2  +\csc^2(\pi t)\pi^2 \geq 0.
\end{align}}
To show $Q(t;\alpha)\geq 0$, we can treat $Q(t;\alpha)$ as a function of $\alpha$ (for fixed $t$). Because both $\frac{1}{\sin(x)}$ and $\frac{1}{\cos(x)}$ are convex functions of $x\in[0,\pi/2]$, we know $Q(t;\alpha)$ is a concave function of $\alpha$ (for fixed $t$). It is easy to check that
{\scriptsize\begin{align}\notag
\underset{\alpha\rightarrow1+}{\lim}{Q(t;\alpha)} = 0, \hspace{0.5in} \underset{\alpha\rightarrow2-}{\lim}{Q(t;\alpha)} = 0.
\end{align}}
Because $Q(t;\alpha)$ is concave in $\alpha\in[1,2]$, we must have $Q(t;\alpha)\geq 0$; and consequently, $W^{\prime\prime}(t) \geq 0$ and $g^\prime(t)\geq 0$. Therefore, we have proved that (\ref{eqn_proof_gm_coefficent}) decreases monotonically with increasing $k$, when $1<\alpha\leq 2$.

For $\alpha<1$ (i.e., $\kappa(\alpha)=\alpha<1$), we prove the monotonicity by a different technique. First, using infinite-product representations \cite[8.322,1.431.1]{Book:Gradshteyn_00},
{\scriptsize\begin{align}\notag
&\Gamma(z) = \frac{\exp\left(-\gamma_ez\right)}{z}\prod_{s=1}^\infty
\left(1+\frac{z}{s}\right)^{-1}\exp\left(\frac{z}{s}\right),\\\notag
&\sin(z) =z\prod_{s=1}^\infty\left(1-\frac{z^2}{s^2\pi^2}\right),
\end{align}}
we can rewrite (\ref{eqn_proof_gm_coefficent}) as
{\scriptsize\begin{align}\notag
&\left[2\cos\left(\frac{\kappa(\alpha)\pi}{2k}\right)\frac{\Gamma\left(\frac{\alpha}{k}\right)\sin\left(\frac{\pi\alpha}{2k}\right)}{\Gamma\left(\frac{1}{k}\right)\sin\left(\frac{\pi}{k}\right)} \right]^k
= \exp\left(-\gamma_e(\alpha-1)\right)\times\\\notag
&\left(
\prod_{s=1}^\infty\exp\left(\frac{\alpha-1}{sk}\right)\left(1+\frac{\alpha}{ks}\right)^{-1}\left(1+\frac{1}{ks}\right)
\left(1-\frac{\alpha^2}{k^2s^2}\right)\left(1-\frac{1}{s^2k^2}\right)^{-1}\right)^k.
\end{align}}
To show its monotonicity, it suffices to show for any $s\geq 1$
{\scriptsize\begin{align}\notag
\left(\left(1+\frac{\alpha}{ks}\right)^{-1}\left(1+\frac{1}{ks}\right)\left(1-\frac{\alpha^2}{k^2s^2}\right)\left(1-\frac{1}{s^2k^2}\right)^{-1}
\right)^k
\end{align}}
decreases monotonically, which is equivalent to show the monotonicity of $g(t)$ with increasing $t$, for $t\geq2$, where
{\scriptsize\begin{align}\notag
g(t) = t\log\left(
\left(1+\frac{\alpha}{t}\right)^{-1}\left(1+\frac{1}{t}\right)\left(1-\frac{\alpha^2}{t^2}\right)\left(1-\frac{1}{t^2}\right)^{-1}
\right) = t \log\left(\frac{t-\alpha}{t-1}\right).
\end{align}}
It is straightforward to show that $t \log\left(\frac{t-\alpha}{t-1}\right)$ is  monotonically decreasing with increasing $t$ ($t\geq 2$), for $\alpha<1$.

To this end, we have proved that for $0<\alpha\leq 2$ ($\alpha \neq 1$),
{\scriptsize\begin{align}\notag
&\left[\cos\left(\frac{\kappa(\alpha)\pi}{2k}\right)\frac{2}{\pi}\Gamma\left(\frac{\alpha}{k}\right)\Gamma\left(1-\frac{1}{k}\right)\sin\left(\frac{\pi}{2}\frac{\alpha}{k}\right)\right]^k
\rightarrow \exp\left(-\gamma_e\left(\alpha-1\right)\right),
\end{align}}
\noindent monotonically with increasing $k$ ($k\geq 2$).

\section{Proof of Lemma \ref{lem_gm_bounds}}\label{proof_lem_gm_bounds}

We first find the constant $G_{R,gm}$ in the right tail bound
{\scriptsize\begin{align}\notag
\mathbf{Pr}\left(\hat{F}_{(\alpha),gm,b} - F_{(\alpha)} \geq \epsilon F_{(\alpha)} \right) \leq
\exp\left(-k\frac{\epsilon^2}{G_{R,gm}}\right),  \ \
\epsilon>0.
\end{align}}
For {\small$0<t<k$}, the Markov moment bound yields
{\scriptsize\begin{align}\notag
&\mathbf{Pr}\left(\hat{F}_{(\alpha),gm,b} - F_{(\alpha)} \geq \epsilon F_{(\alpha)}\right)
\leq \frac{\text{E}\left(\hat{F}_{(\alpha),gm}\right)^t}{(1+\epsilon)^tF_{(\alpha)}^t}\\\notag
=&  \frac{\left[\cos\left(\frac{\kappa(\alpha)\pi}{2k}t\right)\frac{2}{\pi}\Gamma\left(\frac{\alpha t}{k}\right)\Gamma\left(1-\frac{t}{k}\right)\sin\left(\frac{\pi\alpha t}{2k}\right)\right]^k}{(1+\epsilon)^{t}\exp\left(-t\gamma_e (\alpha -1)\right)}.
\end{align}}
\noindent We need to find the $t$ that minimizes the upper bound. For convenience, we consider its logarithm, i.e.,
{\scriptsize\begin{align}\notag
&g(t) = t\gamma_e\left(\alpha-1\right) -t\log(1+\epsilon)+ \\\notag &\hspace{0.5in}k\log\left(\cos\left(\frac{\kappa(\alpha)\pi}{2k}t\right)\frac{2}{\pi}\Gamma\left(\frac{\alpha t}{k}\right)\Gamma\left(1-\frac{t}{k}\right)\sin\left(\frac{\pi\alpha t}{2k}\right)\right),
\end{align}}
\noindent whose first and second derivatives (with respect to $t$) are
{\scriptsize\begin{align}\notag
&g^\prime(t) = \gamma_e (\alpha -1)-\log(1+\epsilon)- \frac{\kappa(\alpha)\pi}{2}{\tan\left(\frac{\kappa(\alpha)\pi}{2k}t\right)} + \frac{\alpha\pi/2}{\tan\left(\frac{\alpha\pi t}{2k}\right)}\\\notag
&\hspace{0.5in}+
\psi\left(\frac{\alpha t}{k}\right)\alpha - \psi\left(1-\frac{t}{k}\right),\\\notag
&tg^{\prime\prime}(t)  =-\left(\frac{\kappa(\alpha)\pi}{2}\right)^2{\sec^2\left(\frac{\kappa(\alpha)\pi}{2k}t\right)}
- \left(\frac{\alpha\pi}{2}\right)^2\csc^2\left(\frac{\alpha\pi t}{2k}\right)\\\notag&\hspace{0.5in}
+\alpha^2\psi^\prime\left(\frac{\alpha t}{k}\right) +  \psi^\prime\left(1-\frac{t}{k}\right).
\end{align}}

We need to show that $g(t)$ is a convex function. By the following expansions:
\cite[1.422.2, 1.422.4, 8.363.8]{Book:Gradshteyn_00}
{\scriptsize\begin{align}\notag
&\sec^2\left(\frac{\pi x}{2}\right) = \frac{4}{\pi^2}\sum_{j=1}^\infty\left(\frac{1}{(2j-1-x)^2}+\frac{1}{(2j-1+x)^2}\right),\\\notag
&\csc^2(\pi x) = \frac{1}{\pi^2x^2} +\frac{2}{\pi^2}\sum_{j=1}^\infty \frac{x^2+j^2}{(x^2-j^2)^2},
\hspace{0.2in} \psi^\prime(x) = \sum_{j=0}^\infty\frac{1}{(x+j)^2},
\end{align}}
we can  rewrite
{\scriptsize\begin{align}\notag
&kg^{\prime\prime}(t) = -\kappa^2\sum_{j=1}^\infty\left(\frac{1}{(2j-1-\kappa t/k)^2}+\frac{1}{(2j-1+\kappa t/k)^2}\right)
- \frac{k^2}{t^2}\\\notag
& - \frac{\alpha^2}{2}\sum_{j=1}^\infty \frac{(\alpha t/2k)^2+j^2}{((\alpha t/2k)^2-j^2)^2}
+\alpha^2\sum_{j=0}^\infty\frac{1}{(\alpha t/k + j)^2} + \sum_{j=0}^\infty \frac{1}{(1-t/k+j)^2}\\\notag
=& -\kappa^2\sum_{j=1}^\infty\left(\frac{1}{(2j-1-\kappa t/k)^2}+\frac{1}{(2j-1+\kappa t/k)^2}\right)\\\notag
&\hspace{0.2in}- \alpha^2\sum_{j=1}^\infty\left(\frac{1}{(\alpha t/k - 2j)^2} + \frac{1}{(\alpha t/k + 2j)^2}\right)\\\notag
&\hspace{0.2in}+\alpha^2\sum_{j=1}^\infty\frac{1}{(\alpha t/k + j)^2} + \sum_{j=1}^\infty \frac{1}{(j-t/k)^2}.
\end{align}}
If $\alpha <1$, i.e., $\kappa(\alpha) = \alpha$, then
{\scriptsize\begin{align}\notag
&kg^{\prime\prime}(t)=- \alpha^2\sum_{j=1}^\infty\left(\frac{1}{(\alpha t/k - j)^2} + \frac{1}{(\alpha t/k + j)^2}\right)
+\alpha^2\sum_{j=1}^\infty\frac{1}{(\alpha t/k + j)^2}\\\notag
& + \sum_{j=1}^\infty \frac{1}{(j-t/k)^2}
=- \alpha^2\sum_{j=1}^\infty\frac{1}{(j-\alpha t/k)^2} + \sum_{j=1}^\infty \frac{1}{(j-t/k)^2}\geq \  0,
\end{align}}
\noindent because $\alpha <1$ and $0<t<k$.

If $\alpha >1$, i.e., $\kappa(\alpha) = 2-\alpha<1$, then
{\scriptsize\begin{align}\notag
&kg^{\prime\prime}(t)
= -\kappa^2\sum_{j=1}^\infty\left(\frac{1}{(2j-1-\kappa t/k)^2}+\frac{1}{(2j-1+\kappa t/k)^2}\right)\\\notag
&- \alpha^2\sum_{j=1}^\infty\left(\frac{1}{(\alpha t/k - 2j)^2} + \frac{1}{(\alpha t/k + 2j)^2}\right)
+\alpha^2\sum_{j=1}^\infty\frac{1}{(\alpha t/k + 2j)^2}\\\notag
& + \alpha^2\sum_{j=1}^\infty\frac{1}{(\alpha t/k + 2j-1)^2}
 + \sum_{j=1}^\infty \frac{1}{(2j-t/k)^2} + \sum_{j=1}^\infty \frac{1}{(2j-1-t/k)^2} \\\notag
\geq& -\kappa^2\sum_{j=1}^\infty\frac{1}{(2j-1+\kappa t/k)^2}
- \alpha^2\sum_{j=1}^\infty\frac{1}{(2j-\alpha t/k )^2}\\\notag
&+ \alpha^2\sum_{j=1}^\infty\frac{1}{(\alpha t/k + 2j-1)^2} + \sum_{j=1}^\infty \frac{1}{(2j-t/k)^2} \\\notag
 =&\left( -\sum_{j=1}^\infty\frac{1}{((2j-1)/\kappa+ t/k)^2}+ \sum_{j=1}^\infty\frac{1}{((2j-1)/\alpha + t/k )^2}\right)\\\notag
&+\left(- \sum_{j=1}^\infty\frac{1}{(2j/\alpha-t/k )^2}
 + \sum_{j=1}^\infty \frac{1}{(2j-t/k)^2}\right)
 \geq 0, \hspace{0.2in} (\text{because} \ \alpha>\kappa).
  \end{align}}

Since we have proved that $g^{\prime\prime}(t)$, i.e., $g(t)$ is a convex function, one can find the optimal $t$ by
solving $g^\prime(t) = 0$:
{\scriptsize\begin{align}\notag
&\gamma_e (\alpha -1)-\log(1+\epsilon)- \frac{\kappa(\alpha)\pi}{2}{\tan\left(\frac{\kappa(\alpha)\pi}{2k}t\right)} + \frac{\alpha\pi/2}{\tan\left(\frac{\alpha\pi t}{2k}\right)}+
\\\notag &\hspace{1.0in} +
\psi\left(\frac{\alpha t}{k}\right)\alpha - \psi\left(1-\frac{t}{k}\right)=0,
\end{align}}
\noindent We let the solution be $t = C_R k$, where $C_R$ is the solution to
{\scriptsize\begin{align}\notag
&\gamma_e (\alpha -1)-\log(1+\epsilon)- \frac{\kappa(\alpha)\pi}{2}{\tan\left(\frac{\kappa(\alpha)\pi}{2}C_R\right)} + \frac{\alpha\pi/2}{\tan\left(\frac{\alpha\pi }{2}C_R\right)}\\\notag&\hspace{1in}+
\psi\left(\alpha C_R\right)\alpha - \psi\left(1-C_R\right)=0.
\end{align}}

Alternatively,  we can seek
a ``sub-optimal'' (but asymptotically optimal) solution using the asymptotic expression for
{\small$\text{E}\left(\hat{F}_{(\alpha),gm}\right)^t$} in
Lemma \ref{lem_gm_moments}, i.e., the $t$ that minimizes
{\scriptsize\begin{align}\notag
(1+\epsilon)^{-t}\exp\left(\frac{1}{k}\frac{\pi^2}{24}\left(t^2-t\right)\left(2+\alpha^2-3\kappa^2(\alpha)\right)\right),
\end{align}}
\noindent whose minimum is attained at
{\scriptsize\begin{align}\notag
t = k \frac{\log(1+\epsilon)}{(2+\alpha^2-3\kappa^2(\alpha))\pi^2/12} + \frac{1}{2}.
\end{align}}
\noindent  This approximation can be useful (e.g.,) for serving the initial guess for $C_R$ in a numerical procedure.

Assume we know $C_R$ (e.g., by a numerical procedure), we can then express the right tail bound as
{\scriptsize\begin{align}\notag
&\mathbf{Pr}\left(\hat{F}_{(\alpha),gm,b} - F_{(\alpha)} \geq \epsilon F_{(\alpha)}\right)
\leq  \exp\left(-k \frac{\epsilon^2}{G_{R,gm}}\right),\\\notag
&\frac{\epsilon^2}{G_{R,gm}} =
C_R \log(1+\epsilon) - C_R \gamma_e(\alpha-1)\\\notag
&\hspace{0.5in} -  \log\left(\cos\left(\frac{\kappa(\alpha)\pi C_R}{2}\right)
\frac{2}{\pi}\Gamma\left(\alpha C_R\right)\Gamma\left(1-C_R\right)\sin\left(\frac{\pi\alpha C_R}{2}\right)\right).
\end{align}}

Next, we find the constant $G_{L,gm}$ in  the left tail  bound
{\scriptsize\begin{align}\notag
\mathbf{Pr}\left(\hat{F}_{(\alpha),gm,b} - F_{(\alpha)} \leq -\epsilon F_{(\alpha)}\right) \leq
\exp\left(-k\frac{\epsilon^2}{G_{L, gm}}\right),  \ \
0<\epsilon < 1.
\end{align}}

From Lemma \ref{lem_gm_moments}, we know that, for any $t$, where $0< t<k/\alpha$ if $\alpha>1$ and $t>0$ if $\alpha<1$,
{\scriptsize\begin{align}\notag
&\mathbf{Pr}\left(\hat{F}_{(\alpha),gm,b} \leq  (1-\epsilon)F_{(\alpha)}
\right)  \\\notag
=& \mathbf{Pr}\left(\hat{F}_{(\alpha),gm,b}^{-t} \geq  (1-\epsilon)^{-t}F_{(\alpha)}^{-t}
\right)
\leq
\frac{\text{E}\left(\hat{F}_{(\alpha),gm,b}^{-t}\right)}{(1-\epsilon)^{-t}F_{(\alpha)}^{-t}}\\\notag
=& (1-\epsilon)^t\frac{\left[ -\cos\left(\frac{\kappa(\alpha)\pi}{2k}t\right) \frac{2}{\pi}\Gamma\left(-\frac{\alpha t}{k}\right)\Gamma\left(1+\frac{t}{k}\right)\sin\left(\frac{\pi\alpha t}{2k}\right)\right]^k}
{\exp\left(t\gamma_e(\alpha-1)\right)}\\\notag
=&(1-\epsilon)^t\exp\left(-t\gamma_e(\alpha-1)\right)\frac{\left[
\cos\left(\frac{\kappa(\alpha)\pi}{2k}t\right)\Gamma\left(1+\frac{t}{k}\right)
\right]^k}{\left[
\Gamma\left(1+\frac{\alpha t}{k}\right)\cos\left(\frac{\pi\alpha t}{2k}\right)
\right]^k}\\\notag
=&(1-\epsilon)^t\exp\left(-t\gamma_e(\alpha-1)\right)\frac{\left[
\cos\left(\frac{\kappa(\alpha)\pi}{2k}t\right)\Gamma\left(\frac{t}{k}\right)
\right]^k}{\left[\alpha
\Gamma\left(\frac{\alpha t}{k}\right)\cos\left(\frac{\pi\alpha t}{2k}\right)
\right]^k}
\end{align}}
\noindent whose minimum is attained at $t = C_L k$ (we skip the proof of convexity) such that
{\scriptsize\begin{align}\notag
&\log(1-\epsilon) - \gamma_e(\alpha-1) - \frac{\kappa(\alpha)\pi}{2} \tan\left(\frac{\kappa(\alpha)\pi}{2}C_L\right) +\frac{\alpha\pi}{2} { \tan\left(\frac{\alpha\pi}{2}C_L\right)}\\\notag
&\hspace{0.5in} -\psi\left(\alpha C_L\right)\alpha+ \psi\left(C_L\right)=0.
\end{align}}
Thus, we show the left tail bound\vspace{-0.05in}
{\scriptsize\begin{align}\notag
&\mathbf{Pr}\left(\hat{F}_{(\alpha),gm,b} - F_{(\alpha)} \leq -\epsilon F_{(\alpha)}\right)\leq \exp\left(-k \frac{\epsilon^2}{G_{L,gm}}\right),\\\notag
&\frac{\epsilon^2}{G_{L,gm}}
=
-C_L \log(1-\epsilon)  + C_L\gamma_e(\alpha-1)+\log\alpha\\\notag
&\hspace{0.1in} -  \log\left(\cos\left(\frac{\kappa(\alpha)\pi}{2}C_L\right)\Gamma\left(C_L\right)\right)+\log\left(   \Gamma\left(\alpha C_L\right)\cos\left(\frac{\pi\alpha C_L}{2}\right)\right).
\end{align}}

\section{Proof of Lemma \ref{lem_G_gm_rate}}\label{proof_lem_G_gm_rate}
First, we consider the right bound. From Lemma \ref{lem_gm_bounds},
{\scriptsize\begin{align}\notag
&\frac{\epsilon^2}{G_{R,gm}} =
C_R \log(1+\epsilon) - C_R \gamma_e(\alpha-1)
 \\\notag
 & - \log\left(\cos\left(\frac{\kappa(\alpha)\pi C_R}{2}\right)
\frac{2}{\pi}\Gamma\left(\alpha C_R\right)\Gamma\left(1-C_R\right)\sin\left(\frac{\pi\alpha C_R}{2}\right)\right),
\end{align}}
\noindent and $C_R$ is the solution to  $g_1(C_R,\alpha,\epsilon) = 0$,
{\scriptsize\begin{align}\notag
&g_1(C_R,\alpha,\epsilon) = -\gamma_e (\alpha -1)+\log(1+\epsilon)+ \frac{\kappa(\alpha)\pi}{2}{\tan\left(\frac{\kappa(\alpha)\pi}{2}C_R\right)}\\\notag
 &\hspace{0.5in}- \frac{\alpha\pi/2}{\tan\left(\frac{\alpha\pi }{2}C_R\right)}-
\psi\left(\alpha C_R\right)\alpha + \psi\left(1-C_R\right)=0.
\end{align}}
Using series representations in \cite[1.421.1,1.421.3,8.362.1]{Book:Gradshteyn_00}
{\scriptsize\begin{align}\notag
&\tan\left(\frac{\pi x}{2}\right) = \frac{4x}{\pi}\sum_{j=1}^\infty \frac{1}{(2j-1)^2-x^2},\hspace{0.05in}
\frac{1}{\tan\left(\pi x\right)} = \frac{1}{\pi x} + \frac{2x}{\pi}\sum_{j=1}^\infty \frac{1}{x^2-j^2},\\\notag
&\psi(x) 
= -\gamma_e - \frac{1}{x} + x\sum_{j=1}^\infty \frac{1}{j(x+j)},
\end{align}}
\noindent we rewrite $g_1$ as
{\scriptsize\begin{align}\notag
&g_1 = -\gamma_e (\alpha - 1) +\log(1+\epsilon)+ \frac{\kappa\pi}{2}\frac{4\kappa C_R}{\pi}\sum_{j=1}^\infty \frac{1}{(2j-1)^2-(\kappa C_R)^2}\\\notag
 &- \frac{\alpha\pi}{2}  \left(\frac{2}{\pi \alpha C_R} + \frac{\alpha C_R}{\pi}\sum_{j=1}^\infty \frac{1}{(\alpha C_R/2)^2-j^2}\right)\\\notag
&-\alpha\left( -\gamma_e -  \frac{1}{\alpha C_R} + \alpha C_R\sum_{j=1}^\infty\frac{1}{j(\alpha C_R+j)}\right)\\\notag
&+\left(-\gamma_e - \frac{1}{1-C_R} + (1-C_R) \sum_{j=1}^\infty\frac{1}{j(1-C_R+j)}\right)
\\\notag
=& \log(1+\epsilon)+ 2\kappa^2 C_R\sum_{j=1}^\infty \frac{1}{(2j-1)^2-(\kappa C_R)^2} + 2\alpha^2C_R  \sum_{j=1}^\infty \frac{1}{(2j)^2 - (\alpha C_R)^2}\\\notag
&-\alpha^2C_R\sum_{j=1}^\infty\frac{1}{j(\alpha C_R+j)}
+ (1-C_R)\sum_{j=1}^\infty\frac{1}{j(1-C_R+j)} -\frac{1}{1-C_R}\\\notag
=& \log(1+\epsilon)+ \kappa\sum_{j=1}^\infty \left(\frac{1}{2j+1-\kappa C_R} - \frac{1}{2j-1+\kappa C_R}\right)\\\notag
&+ \alpha\sum_{j=1}^\infty \left(\frac{1}{2j - \alpha C_R}-\frac{1}{2j+\alpha C_R}\right)
-\alpha\sum_{j=1}^\infty\left(\frac{1}{j} - \frac{1}{\alpha C_R+j}\right)\\\notag
&+ \sum_{j=1}^\infty\left(\frac{1}{j}-\frac{1}{1-C_R+j}\right) + \frac{\kappa}{1-\kappa C_R}-\frac{1}{1-C_R}
\end{align}}
We show that, as $\alpha\rightarrow 1$, i.e., $\kappa\rightarrow 1$, the term
{\scriptsize\begin{align}\notag
\lim_{\alpha\rightarrow 1}& \ \  \kappa\sum_{j=1}^\infty \left(\frac{1}{2j+1-\kappa C_R} - \frac{1}{2j-1+\kappa C_R}\right)\\\notag
 &+ \alpha\sum_{j=1}^\infty \left(\frac{1}{2j - \alpha C_R}-\frac{1}{2j+\alpha C_R}\right)\\\notag
& -\alpha\sum_{j=1}^\infty\left(\frac{1}{j} - \frac{1}{\alpha C_R+j}\right)
+ \sum_{j=1}^\infty\left(\frac{1}{j}-\frac{1}{1-C_R+j}\right)\\\notag
=\lim_{\alpha\rightarrow 1}& \ \  \sum_{j=1}^\infty \left(\frac{\kappa}{2j+1-\kappa C_R} + \frac{\alpha}{2j - \alpha C_R}\right)\\\notag
& - \sum_{j=1}^\infty\left( \frac{\kappa}{2j-1+\kappa C_R} +\frac{\alpha}{2j+\alpha C_R}\right)\\\notag
& -\alpha\sum_{j=1}^\infty\left(\frac{1}{j} - \frac{1}{\alpha C_R+j}\right)
+ \sum_{j=1}^\infty\left(\frac{1}{j}-\frac{1}{1-C_R+j}\right)\\\notag
=\lim_{\alpha\rightarrow 1}& \ \  \sum_{j=1}^\infty\frac{\kappa}{1+j-\kappa C_R}  - \sum_{j=1}^\infty \frac{\kappa}{j+\kappa C_R} -\alpha\sum_{j=1}^\infty\left(\frac{1}{j} - \frac{1}{\alpha C_R+j}\right)\\\notag
&+ \sum_{j=1}^\infty\left(\frac{1}{j}-\frac{1}{1-C_R+j}\right)=0.
\end{align}}
From Lemma \ref{lem_gm_bounds}, we know  $g_1 = 0$ has a unique well-defined solution for $C_R\in(0,1)$. We need to analyze this term
{\scriptsize\begin{align}\notag
\frac{\kappa}{1-\kappa C_R}-\frac{1}{1-C_R} = \frac{\kappa - 1}{(1-\kappa C_R)(1-C_R)} =\frac{-\Delta}{(1-\kappa C_R)(1-C_R)},
\end{align}}
\noindent which, when $\alpha \rightarrow 1$ (i.e., $\kappa \rightarrow 1$), must approach a finite limit. In other words, $C_R \rightarrow 1$, at the rate $O\left(\sqrt{\Delta}\right)$, i.e.,
{\scriptsize\begin{align}\notag
C_R =  1 - \sqrt{\frac{\Delta}{\log (1+\epsilon)}} + o\left(\sqrt{\Delta}\right).
\end{align}}

By Euler'r reflection formula and series representations,
{\scriptsize\begin{align}\notag
\frac{\epsilon^2}{G_{R,gm}}
=&C_R \log(1+\epsilon) - C_R \gamma_e(\alpha-1)
 + \log\left(\frac{\cos\left(\frac{\alpha\pi C_R}{2}\right) \Gamma(1-\alpha C_R)}{\cos\left(\frac{\kappa\pi C_R}{2}\right)\Gamma(1-C_R)} \right),
\end{align}}
{\scriptsize\begin{align}\notag
&\frac{\cos\left(\frac{\alpha\pi C_R}{2}\right) \Gamma(1-\alpha C_R)}{\cos\left(\frac{\kappa\pi C_R}{2}\right)\Gamma(1-C_R)}\\\notag
=&\exp(\gamma_e(\alpha-1) C_R)\frac{1- C_R}{1-\alpha C_R}
\prod_{j=0}^\infty \left(1-\frac{\alpha^2 C_R^2}{(2j+1)^2}\right) \left(1-\frac{\kappa^2 C_R^2}{(2j+1)^2}\right)^{-1} \\\notag &\times\prod_{j=1}^\infty\exp\left(\frac{(1-\alpha) C_R}{j}\right) \left(1+\frac{1- C_R}{j}\right)\left(1+\frac{1-\alpha C_R}{j}\right)^{-1} \\\notag
=&\exp(\gamma_e(\alpha -1) C_R)\frac{(1+\alpha C_R)(1- C_R)}{1-\kappa^2C_R^2} \prod_{j=1}^\infty \left(1-\frac{\alpha^2 C_R^2}{(2j+1)^2}\right)\\\notag
&\times \left(1-\frac{\kappa^2 C_R^2}{(2j+1)^2}\right)^{-1}
\exp\left(\frac{(1-\alpha) C_R}{j}\right) \left(1+\frac{1- C_R}{j}\right)\left(1+\frac{1-\alpha C_R}{j}\right)^{-1},
\end{align}}
taking logarithm of which yields
{\scriptsize\begin{align}\notag
&\log \frac{\cos\left(\frac{\alpha\pi C_R}{2}\right) \Gamma(1-\alpha C_R)}{\cos\left(\frac{\kappa\pi C_R}{2}\right)\Gamma(1-C_R)}
=\gamma_e(\alpha-1) C_R + \log \frac{(1+\alpha C_R)(1- C_R)}{1-\kappa^2C_R^2}\\\notag
&\hspace{0.5in}+\sum_{j=1}^\infty \log \frac{\left(1-\frac{\alpha^2 C_R^2}{(2j+1)^2}\right)}{\left(1-\frac{\kappa^2 C_R^2}{(2j+1)^2}\right)} + \left(\frac{(1-\alpha) C_R}{j}\right) + \log \frac{ \left(1+\frac{1- C_R}{j}\right)}{\left(1+\frac{1-\alpha C_R}{j}\right)}.
\end{align}}
If $\alpha <1$, i.e., $\kappa =\alpha = 1-\Delta$, then
{\scriptsize\begin{align}\notag
&\log \frac{\cos\left(\frac{\alpha\pi C_R}{2}\right) \Gamma(1-\alpha C_R)}{\cos\left(\frac{\kappa\pi C_R}{2}\right)\Gamma(1-C_R)}\\\notag
=&-\gamma_e\Delta C_R + \log \frac{1- C_R}{1-\alpha C_R}+\sum_{j=1}^\infty \left(\frac{(1-\alpha) C_R}{j}\right) + \log \frac{ \left(1+\frac{1- C_R}{j}\right)}{\left(1+\frac{1-\alpha C_R}{j}\right)}\\\notag
=&-\gamma_e\Delta C_R - \log\left(1+ \frac{\Delta C_R}{1-C_R}\right)+\sum_{j=1}^\infty \frac{1}{2}\left(\frac{1-\alpha C_R}{j}\right)^2 - \frac{1}{2}\left(\frac{1-C_R}{j}\right)^2 ...\\\notag
=&-\gamma_e\Delta C_R - \log\left(1+ \frac{\Delta C_R}{1-C_R}\right)+ \frac{\pi^2}{12}C_R\Delta (2-\alpha C_R - C_R) +...
\end{align}}
Thus, for $\alpha <1$, as {\scriptsize$C_R = 1-\sqrt{\frac{\Delta}{\log (1+\epsilon)}} + o\left(\sqrt{\Delta}\right)$}, we obtain
{\scriptsize\begin{align}\notag
\frac{\epsilon^2}{G_{R,gm}}  =& C_R \log(1+\epsilon) - \frac{\Delta C_R}{1-C_R} + \frac{\pi^2}{12}C_R\Delta (2-\alpha C_R - C_R) +... \\\notag
=& \log(1+\epsilon) - 2\sqrt{\Delta\log\left(1+\epsilon\right)}+o\left(\sqrt{\Delta}\right)
\end{align}}

If $\alpha >1$, i.e., $\alpha = 1+\Delta$ and $\kappa = 1-\Delta$, then
{\scriptsize\begin{align}\notag
&\log \frac{\cos\left(\frac{\alpha\pi C_R}{2}\right) \Gamma(1-\alpha C_R)}{\cos\left(\frac{\kappa\pi C_R}{2}\right)\Gamma(1-C_R)}\\\notag
=&\gamma_e\Delta C_R + \log \frac{(1+\alpha C_R)(1- C_R)}{1-\kappa^2C_R^2}
+\sum_{j=1}^\infty \log \frac{\left(1-\frac{\alpha^2 C_R^2}{(2j+1)^2}\right)}{\left(1-\frac{\kappa^2 C_R^2}{(2j+1)^2}\right)} +...
\end{align}}
Also
{\scriptsize\begin{align}\notag
&\log \frac{(1+\alpha C_R)(1- C_R)}{1-\kappa^2C_R^2} = \log \frac{1+\alpha C_R}{1+\kappa C_R} - \log \frac{1-\kappa C_R}{1-C_R}\\\notag
=&\log\left(1+ \frac{2\Delta C_R}{1+\kappa C_R}\right)  - \log\left(1+\frac{\Delta C_R}{1-C_R}\right)\\\notag
=& -\sqrt{\Delta\log(1+\epsilon)} + o\left(\sqrt{\Delta}\right),
\end{align}}
and
{\scriptsize\begin{align}\notag
&\sum_{j=1}^\infty \log \frac{\left(1-\frac{\alpha^2 C_R^2}{(2j+1)^2}\right)}{\left(1-\frac{\kappa^2 C_R^2}{(2j+1)^2}\right)} =\sum_{j=1}^\infty  \log \frac {1+\frac{\alpha C_R}{2j+1}}{1+ \frac{\kappa C_R}{2j+1}} + \log \frac {1-\frac{\alpha C_R}{2j+1}}{1-\frac{\kappa C_R}{2j+1}} \\\notag
=&\sum_{j=1}^\infty  \log \left(1+\frac {\frac{2\Delta C_R}{2j+1}}{1+ \frac{\kappa C_R}{2j+1}}\right) + \log \left(1-\frac {\frac{2\Delta C_R}{2j+1}}{1-\frac{\kappa C_R}{2j+1}}\right) = O\left(\Delta\right).
 \end{align}}
Therefore, for $\alpha >1$, we also have
{\scriptsize\begin{align}\notag
\frac{\epsilon^2}{G_{R,gm}}
=& \log(1+\epsilon) - 2\sqrt{\Delta\log\left(1+\epsilon\right)}+o\left(\sqrt{\Delta}\right).
\end{align}}
In other words, as $\alpha \rightarrow 1$, the constant $G_{R,gm}$ converges to $\frac{\epsilon^2}{\log(1+\epsilon)}$ at the rate $O\left(\sqrt{\Delta}\right)$. \\

Next, we consider the left bound. From Lemma \ref{lem_gm_bounds},
{\scriptsize\begin{align}\notag
&\mathbf{Pr}\left(\hat{F}_{(\alpha),gm,b} - F_{(\alpha)} \leq -\epsilon F_{(\alpha)}\right)\leq \exp\left(-k \frac{\epsilon^2}{G_{L,gm}}\right),
\end{align}}
\noindent where
{\scriptsize\begin{align}\notag
&\frac{\epsilon^2}{G_{L,gm}}
=
-C_L \log(1-\epsilon)  + C_L\gamma_e(\alpha-1)+\log\alpha\\\notag
&\hspace{0.in} -  \log\left(\cos\left(\frac{\kappa(\alpha)\pi}{2}C_L\right)\Gamma\left(C_L\right)\right)+\log\left(   \Gamma\left(\alpha C_L\right)\cos\left(\frac{\pi\alpha C_L}{2}\right)\right).
\end{align}}
\noindent and $C_L$ is the solution to  $g_2(C_L,\alpha,\epsilon) = 0$,
{\scriptsize\begin{align}\notag
&g_2(C_L,\alpha,\epsilon) =\log(1-\epsilon) - \gamma_e(\alpha-1) - \frac{\kappa(\alpha)\pi}{2} \tan\left(\frac{\kappa(\alpha)\pi}{2}C_L\right)\\\notag
&\hspace{0.5in} +\frac{\alpha\pi}{2} { \tan\left(\frac{\alpha\pi}{2}C_L\right)}
 -\psi\left(\alpha C_L\right)\alpha+ \psi\left(C_L\right)=0.
\end{align}}

Using series representations,  we rewrite $g_2$ as
{\scriptsize\begin{align}\notag
&g_2 = -\gamma_e (\alpha - 1) +\log(1-\epsilon)- \frac{\kappa\pi}{2}\frac{4\kappa C_L}{\pi}\sum_{j=1}^\infty \frac{1}{(2j-1)^2-(\kappa C_L)^2}\\\notag
 &+ \frac{\alpha\pi}{2}\frac{4\alpha C_L}{\pi}\sum_{j=1}^\infty \frac{1}{(2j-1)^2-(\alpha C_L)^2}
 \\\notag
&-\alpha\left( -\gamma_e -  \frac{1}{\alpha C_L}+ (\alpha C_L)\sum_{j=1}^\infty\frac{1}{j(\alpha C_L+j)}\right)\\\notag
&+\left(-\gamma_e - \frac{1}{C_L} + C_L \sum_{j=1}^\infty\frac{1}{j(C_L+j)}\right)
\\\notag
=&- 2\kappa^2 C_L\sum_{j=1}^\infty \frac{1}{(2j-1)^2-(\kappa C_L)^2} +
 2\alpha^2 C_L\sum_{j=1}^\infty \frac{1}{(2j-1)^2-(\alpha C_L)^2}\\\notag
&-\alpha^2 C_L\sum_{j=1}^\infty\frac{1}{j(\alpha C_L+j)}
+ C_L\sum_{j=1}^\infty\frac{1}{j(C_L+j)}+ \log(1-\epsilon)\\\notag
=& \log(1-\epsilon)- \kappa\sum_{j=1}^\infty \left(\frac{1}{2j-1-\kappa C_L} - \frac{1}{2j-1+\kappa C_L}\right)+\\\notag
&\alpha\sum_{j=1}^\infty \frac{1}{2j-1 - \alpha C_L}-\frac{1}{2j-1+\alpha C_L}
+(1-\alpha) C_L\sum_{j=1}^\infty\frac{\alpha C_L + j(1+\alpha)}{j(\alpha C_L + j)(C_L+j)}.
\end{align}}

We first consider $\alpha=1+\Delta>1$. In order for $g_2 = 0$ to have a meaningful solution, we must make sure that
{\scriptsize\begin{align}\notag
\frac{-\kappa}{1-\kappa C_L} + \frac{\alpha}{1-\alpha C_L} = \frac{2\Delta}{(1-\kappa C_L)(1-\alpha C_L)} = \frac{2\Delta}{1-2C_L +C_L^2-\Delta^2C_L^2}
\end{align}}
converges to a finite value as $\alpha\rightarrow 1$, i.e., $C_L\rightarrow 1$ also. This provides an approximate solution for $C_L$ when $\alpha>1$:
{\scriptsize\begin{align}\notag
C_L = 1 - \sqrt{\frac{2\Delta}{-\log(1-\epsilon)}} + o\left(\sqrt{\Delta}\right).
\end{align}}
Using series representations, we obtain
{\scriptsize\begin{align}\notag
&C_L\gamma_e(\alpha-1)+\log\alpha+\log\frac{ \Gamma\left(\alpha C_L\right)\cos\left(\frac{\pi\alpha C_L}{2}\right)}{
\cos\left(\frac{\kappa(\alpha)\pi}{2}C_L\right)\Gamma\left(C_L\right)}\\\notag
=&\log\left( \prod_{s=1}^\infty\frac{1+\frac{C_L}{s}}{1+\frac{\alpha C_L}{s}}\exp\left(\frac{\Delta C_L}{s}\right)\prod_{s=0}^\infty\frac{1-\frac{\alpha^2C_L^2}{(2s+1)^2}}{1-\frac{\kappa^2C_L^2}{(2s+1)^2}}\right)\\\notag
=&\sum_{s=1}^\infty\left(-\frac{\Delta C_L}{s+C_L}+\frac{\Delta C_L}{s}+o\left(\Delta\right)\right) + \log\left(\frac{1-\alpha^2C_L^2}{1-\kappa^2C_L^2}\right) +\sum_{s=1}^\infty\log \frac{1-\frac{\alpha^2C_L^2}{(2s+1)^2}}{1-\frac{\kappa^2C_L^2}{(2s+1)^2}}\\\notag
=&-\sqrt{-2\Delta\log(1-\epsilon)} + O\left(\Delta\right).
\end{align}}
Therefore,  for $\alpha >1$
{\scriptsize\begin{align}\notag
G_{L,gm} = \frac{\epsilon^2}{ -\log(1-\epsilon) - 2\sqrt{-2\Delta\log(1-\epsilon)} + o\left(\sqrt{\Delta}\right)}.
\end{align}}

Finally, we need to consider $\alpha <1$. In this case,
{\scriptsize\begin{align}\notag
g_2 =& \log(1-\epsilon)+\Delta C_L\sum_{j=1}^\infty\frac{\alpha C_L + j(1+\alpha)}{j(\alpha C_L + j)(C_L+j)}\\\notag
=&\log(1-\epsilon) + \Delta C_L\left(\sum_{j=1}^\infty\frac{1}{j(j+C_L)} + \sum_{j=1}^\infty\frac{1}{(1+C_L)^2}\right)+o\left(\Delta\right).
\end{align}}
Using properties of Riemann's Zeta function and Bernoulli numbers\cite[9.511,9.521.1,9.61]{Book:Gradshteyn_00}
{\scriptsize\begin{align}\notag
&\sum_{j=1}^\infty\frac{1}{(j+C_L)^2} = -\frac{1}{C_L^2} + \int_{0}^\infty\frac{t\exp(-C_L t)}{1-\exp(-t)}dt\\\notag
=&-\frac{1}{C_L^2} + \int_0^\infty\left(1+\frac{t}{2} + \frac{t^2}{12}+...\right) \exp(-C_L t) dt = \frac{1}{C_L} + O\left(\frac{1}{C_L^2}\right).
\end{align}}
Using the integral relation\cite[0.244.1]{Book:Gradshteyn_00} and treating $C_L$ as a positive integer (which does not affect the asymptotics)
{\scriptsize\begin{align}\notag
&\sum_{j=1}^\infty\frac{1}{j(j+C_L)} = \frac{1}{C_L}\int_0^1\frac{1-t^{C_L}}{1-t}dt\\\notag
=&\frac{1}{C_L}\int_0^1 t^{C_L-1} + t^{C_L-2} + ...+ 1 dt \\\notag
 =&\frac{1}{C_L} \sum_{j=1}^{C_L} \frac{1}{j}= \frac{1}{C_L}\left(\gamma_e + \log C_L+O\left(C_L^{-1}\right)\right).
\end{align}}
Thus, the solution to $g_2 =0$ can be approximated by
{\scriptsize\begin{align}\notag
&\log(1-\epsilon) + \Delta \left(1+\gamma_e + \log C_L\right) +o(\Delta)= 0.
\end{align}}
Again, using series representations, we obtain
{\scriptsize\begin{align}\notag
&C_L\gamma_e(\alpha-1)+\log\alpha+\log\frac{ \Gamma\left(\alpha C_L\right)}{
\Gamma\left(C_L\right)}\\\notag
=&\log\left( \prod_{j=1}^\infty\frac{1+\frac{C_L}{j}}{1+\frac{\alpha C_L}{j}}\exp\left(-\frac{\Delta C_L}{j}\right)\right)\\\notag
=&\sum_{j=1}^\infty\left(\frac{\Delta C_L}{j+C_L}-\frac{\Delta C_L}{j}+...\right)\\\notag
=&-\Delta C_L\left(\gamma_e + \log C_L\right) +...
\end{align}}
Combining the results, we obtain, when $\alpha<1$ and $\Delta\rightarrow 0$,
{\scriptsize\begin{align}\notag
G_{L,gm} = \frac{\epsilon^2}{ \Delta\left(\exp\left(\frac{-\log(1-\epsilon)}{\Delta} -1 - \gamma_e\right)\right)+o\left(\Delta\exp\left(\frac{1}{\Delta}\right)\right)}.
\end{align}}

\vspace{-0.1in}
\section{Proof of Lemma \ref{lem_hm}}\label{proof_lem_hm}

Assume $k$ i.i.d. samples $x_j \sim S(\alpha<1, \beta =1, F_{(\alpha)})$. Using the $(-\alpha)$th moment in Lemma \ref{lem_moments} suggests that
{\scriptsize\begin{align}\notag
\hat{R}_{(\alpha)} = \frac{\frac{1}{k}\sum_{j=1}^k|x_j|^{-\alpha}}{\frac{\cos\left(\frac{\alpha\pi}{2}\right)}{\Gamma(1+\alpha)}  },
\end{align}}
\noindent is an unbiased estimator of $d^{-1}_{(\alpha)}$,whose variance is
{\scriptsize\begin{align}\notag
\text{Var}\left(\hat{R}_{(\alpha)}\right) = \frac{d^{-2}_{(\alpha)}}{k}\left(\frac{2\Gamma^2(1+\alpha)}{\Gamma(1+2\alpha)}-1\right).
\end{align}}

We can then estimate $F_{(\alpha)}$ by $\frac{1}{\hat{R}_{(\alpha)}}$, i.e.,
{\scriptsize\begin{align}\notag
\hat{F}_{(\alpha),hm} = \frac{1}{\hat{R}_{(\alpha)}} = \frac{k\frac{\cos\left(\frac{\alpha\pi}{2}\right)}{\Gamma(1+\alpha)}}{\sum_{j=1}^k|x_j|^{-\alpha}}.
\end{align}}
which is biased at the order $O\left(\frac{1}{k}\right)$. To remove the $O\left(\frac{1}{k}\right)$ term of the bias, we recommend a bias-corrected version obtained by Taylor expansions \cite[Theorem 6.1.1]{Book:Lehmann_Casella}:
{\scriptsize\begin{align}\notag
\frac{1}{\hat{R}_{(\alpha)}} -
\frac{\text{Var}\left(\hat{R}_{(\alpha)}\right)}{2}
\left(\frac{2}{F_{(\alpha)}^{-3}}\right),
\end{align}}
from which we obtain the bias-corrected estimator
{\scriptsize\begin{align}\notag
\hat{F}_{(\alpha),hm,c} = \frac{k\frac{\cos\left(\frac{\alpha\pi}{2}\right)}{\Gamma(1+\alpha)}}{\sum_{j=1}^k|x_j|^{-\alpha}}
\left(1- \frac{1}{k}\left(\frac{2\Gamma^2(1+\alpha)}{\Gamma(1+2\alpha)}-1\right) \right),
\end{align}}
whose bias and variance are
{\scriptsize\begin{align}\notag
&\text{E}\left(\hat{F}_{(\alpha),hm,c}\right) = F_{(\alpha)}+O\left(\frac{1}{k^2}\right),\\\notag
&\text{Var}\left(\hat{F}_{(\alpha),hm,c}\right) = \frac{F^{2}_{(\alpha)}}{k}\left(\frac{2\Gamma^2(1+\alpha)}{\Gamma(1+2\alpha)}-1\right) + O\left(\frac{1}{k^2}\right).
\end{align}}

We now study the tail bounds. For convenience, we provide tail bounds for $\hat{F}_{(\alpha),hm}$ instead of $\hat{F}_{(\alpha),hm,c}$. We first analyze the following moment generating function:
{\scriptsize\begin{align}\notag
&\text{E}\left(\exp\left(\frac{F_{(\alpha)}|x_j|^{-\alpha}}{\cos\left(\alpha\pi/2\right)/\Gamma(1+\alpha)}t\right)\right)\\\notag
=& 1+\sum_{m=1}^\infty \frac{t^m}{m!} \text{E}\left( F_{(\alpha)}\left(\frac{|x_j|^{-\alpha}}{\cos\left(\alpha\pi/2\right)/\Gamma(1+\alpha)}\right)^m   \right)\\\notag
=&1+\sum_{m=1}^\infty \frac{t^m}{m!}\frac{\Gamma(1+m)\Gamma^m(1+\alpha)}{\Gamma(1+m\alpha)}
= \sum_{m=0}^\infty \frac{\Gamma^m(1+\alpha)}{\Gamma(1+m\alpha)}t^m.
\end{align}}

For the right tail bound,
{\scriptsize\begin{align}\notag
&\mathbf{Pr}\left(
\hat{F}_{(\alpha),hm} -  F_{(\alpha)} \geq \epsilon F_{(\alpha)}\right) \\\notag
=& \mathbf{Pr}\left( \frac{k\frac{\cos\left(\frac{\alpha\pi}{2}\right)}{\Gamma(1+\alpha)}}{\sum_{j=1}^k|x_j|^{-\alpha}} \geq (1+\epsilon)F_{(\alpha)}\right)\\\notag
=&\mathbf{Pr}\left(\exp\left(-t\left(
\frac{\sum_{j=1}^kF_{(\alpha)}|x_j|^{-\alpha}}{
\cos\left(\alpha\pi/2\right)/\Gamma(1+\alpha)}\right)\right)\geq \exp\left(-t\frac{k}{(1+\epsilon)}\right)\right)\hspace{0.05in} (t>0)\\\notag
\leq&\left(\sum_{m=0}^\infty \frac{\Gamma^m(1+\alpha)}{\Gamma(1+m\alpha)}(-t)^m\right)^k \exp\left(t\frac{k}{(1+\epsilon)}\right)\\\notag
=&\exp\left(-k\left(-\log   \left(\sum_{m=0}^\infty \frac{\Gamma^m(1+\alpha)}{\Gamma(1+m\alpha)}(-t_1^*)^m\right)
-\frac{t_1^*}{1+\epsilon}\right)\right)\\\notag
=&\exp\left(-k\frac{\epsilon^2}{G_{R,hm}}\right),
\end{align}}
\noindent where $t_1^*$ is the solution to
{\scriptsize\begin{align}\notag
\frac{\sum_{m=1}^\infty(-1)^m m (t_1^*)^{m-1}\frac{\Gamma^m(1+\alpha)}{\Gamma(1+m\alpha)} }{\sum_{m=0}^\infty(-1)^m (t_1^*)^{m}\frac{\Gamma^m(1+\alpha)}{\Gamma(1+m\alpha)} } + \frac{1}{1+\epsilon} = 0
.
\end{align}}

For the left tail bound,
{\scriptsize\begin{align}\notag
&\mathbf{Pr}\left(
\hat{F}_{(\alpha),hm} -  F_{(\alpha)} \leq -\epsilon F_{(\alpha)}\right) \\\notag
=& \mathbf{Pr}\left( \frac{k\frac{\cos\left(\frac{\alpha\pi}{2}\right)}{\Gamma(1+\alpha)}}{\sum_{j=1}^k|x_j|^{-\alpha}} \leq (1-\epsilon)F_{(\alpha)}\right)\\\notag
=&\mathbf{Pr}\left(\exp\left(t\left(
\frac{\sum_{j=1}^kF_{(\alpha)}|x_j|^{-\alpha}}{
\cos\left(\alpha\pi/2\right)/\Gamma(1+\alpha)}\right)\right)\geq \exp\left(t\frac{k}{(1-\epsilon)}\right)\right)\hspace{0.2in} (t>0)\\\notag
\leq&\left(\sum_{m=0}^\infty \frac{\Gamma^m(1+\alpha)}{\Gamma(1+m\alpha)}t^m\right)^k \exp\left(-t\frac{k}{(1-\epsilon)}\right)\\\notag
=&\exp\left(-k\left(-\log   \left(\sum_{m=0}^\infty \frac{\Gamma^m(1+\alpha)}{\Gamma(1+m\alpha)}(t_2^*)^m\right)
+\frac{t_2^*}{1-\epsilon}\right)\right),
\end{align}}
\noindent where $t_2^*$ is the solution to
{\scriptsize\begin{align}\notag
\sum_{m=1}^\infty  \left\{\frac{(t_2^*)^{m-1}\Gamma^{m-1}(1+\alpha)}{\Gamma(1+(m-1)\alpha)} - m(1-\epsilon)
\frac{ (t_2^*)^{m-1}\Gamma^m(1+\alpha)}{\Gamma(1+m\alpha)}\right\}=0.
\end{align}}

{\scriptsize\small

}

\end{document}